\documentclass[10pt,twocolumn,letterpaper]{article}

\usepackage{cvpr}              

\usepackage{subcaption}

\usepackage[utf8]{inputenc}
\usepackage[T1]{fontenc}
\usepackage{tabularx}

\usepackage{pgfplots}
\pgfplotsset{compat=1.18}

\usepackage{pifont}

\newcommand{\cmark}{\textcolor{ForestGreen}{\ding{51}}}
\newcommand{\xmark}{\textcolor{Maroon}{\ding{55}}}

\usepackage{booktabs}                    
\usepackage{multirow}                    
\usepackage{makecell}                    
\usepackage{array}                       
\usepackage{float}                       
\usepackage[flushleft]{threeparttable}   
\usepackage{url}
\usepackage[table]{xcolor}
\definecolor{Stripe}{RGB}{248,250,252}          
\definecolor{HeadGradTop}{RGB}{240,245,255}     
\definecolor{HeadGradBot}{RGB}{225,235,250}     

\usepackage{etoolbox}

\newcommand{\best}[1]{\textbf{#1}}

\renewcommand{\arraystretch}{1.08} 

\usepackage{graphicx}
\usepackage{subcaption}

\usepackage{siunitx}

\usepackage[none]{hyphenat}      
\usepackage{textcomp}
\usepackage{url}
\usepackage{verbatim}
\usepackage{pifont}
\usepackage{svg}
\usepackage{stfloats}            

\definecolor{cvprblue}{rgb}{0.21,0.49,0.74}
\usepackage[pagebackref,breaklinks,colorlinks,allcolors=cvprblue]{hyperref}
\usepackage[accsupp]{axessibility}

\def\paperID{26705} 
\def\confName{CVPR}
\def\confYear{2026}

\title{RAVEN: Radar Adaptive Vision Encoders for Efficient Chirp-wise Object Detection and Segmentation}

\author{
Anuvab Sen, Mir Sayeed Mohammad, Saibal Mukhopadhyay\\
Georgia Institute of Technology, Atlanta, Georgia, USA\\
{\tt\small asen74@gatech.edu, mirsayeedmohammad@gatech.edu, saibal.mukhopadhyay@ece.gatech.edu}
}

\begin{document}
\maketitle

\begin{abstract}
We introduce \textbf{RAVEN}, a deep learning  architecture for processing frequency-modulated continuous-wave (FMCW) radar data that is designed for high computational efficiency. RAVEN reduces computation by using a 
learnable antenna mixer module on independent receiver state space encoders (SSM) to compress the virtual MIMO array into a compact set of learned features and by performing per-chirp inference with a calibrated early-exit rule, so the model reaches a decision using only a subset of chirps in a radar frame. These design choices yield up to \textbf{170×} lower computation and \textbf{4×} lower end-to-end latency than conventional frame-based radar backbones, while achieving state-of-the-art detection and BEV free-space segmentation performance on automotive radar datasets.









\end{abstract}   
\section{Introduction} \label{sec:intro}

Millimeter–wave radars are increasingly central to perception on autonomous ground and aerial platforms. Compared to cameras and LiDAR, they remain robust under adverse weather/lighting and directly sense relative velocity via Doppler. These qualities, combined with radar’s lower size, weight, and power profile, make it an attractive sensing modality for mobile platforms \cite{han20234d, fan20244d}. Recent 4D imaging radars further extend range and reliability under fog and rain, offering stronger velocity tracking \cite{zhang2023perception, paek2022k, burnett2022we}. Yet higher spatial and Doppler resolution comes with a steep cost: data volume and compute scale rapidly with antenna count and bandwidth \cite{rebut2022radial, wang2021rodnet, dalbah2024transradar}, making them infeasible for embedded platforms and high-speed applications \cite{patole2017automotive, sun2020mimo, han20234d}. 

Most existing deep learning pipelines for radar perception follow a frame-based paradigm. They first collect all ADC samples for an entire radar frame, apply a sequence of fast Fourier transforms (FFTs) along range, angle, and Doppler dimensions to construct high-resolution range–angle–Doppler (RAD) tensors, and then run dense convolutional or transformer backbones on these tensors. This design exposes rich spatial–Doppler structure, but it also fixes latency to at least one frame interval and requires expensive dense processing on large 3D feature maps, which is problematic on resource-constrained platforms.

\begin{figure*}[htb]
    \centering
    \includegraphics[width = \linewidth]{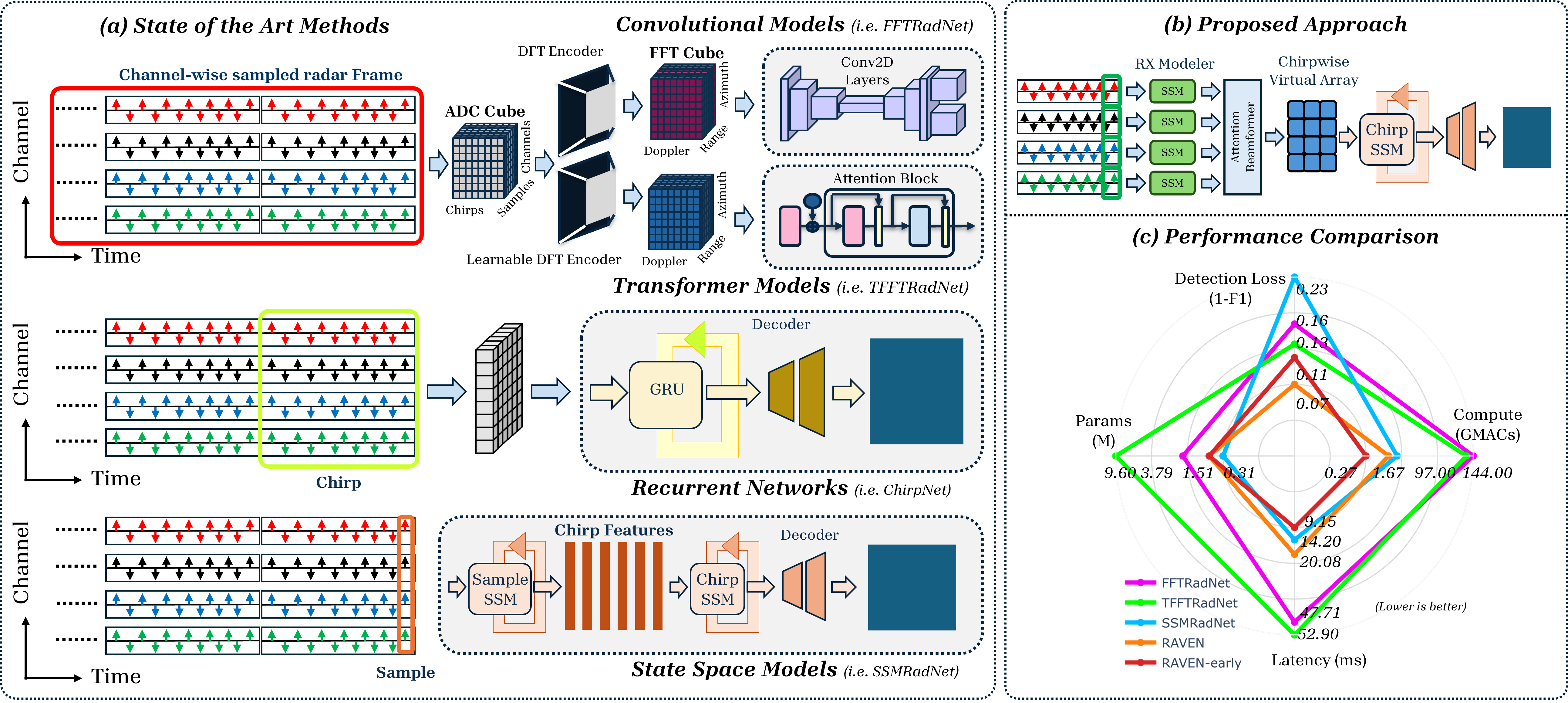}
    \caption{(a) Comparison of traditional radar processing paradigms: frame-wise CNN encoders, chirp-wise recurrent models, and sample-wise streaming SSM pipelines. (b) Our spatial-aware hybrid architecture preserves per-RX structure, performs cross-antenna attention and extracts chirp-wise virtual-array features for lightweight detection. (c) Runtime–performance characterization showing our method achieves higher accuracy at significantly lower latency and compute compared to existing radar perception models \cite{rebut2022radial, giroux2023tfftradnet, sharma2024chirpnet, sen2025ssmradnetsamplewisestatespace}.}
    \label{fig:motivation}
\end{figure*}

\begin{figure}[t]
    \centering
    \includegraphics[width = \linewidth]{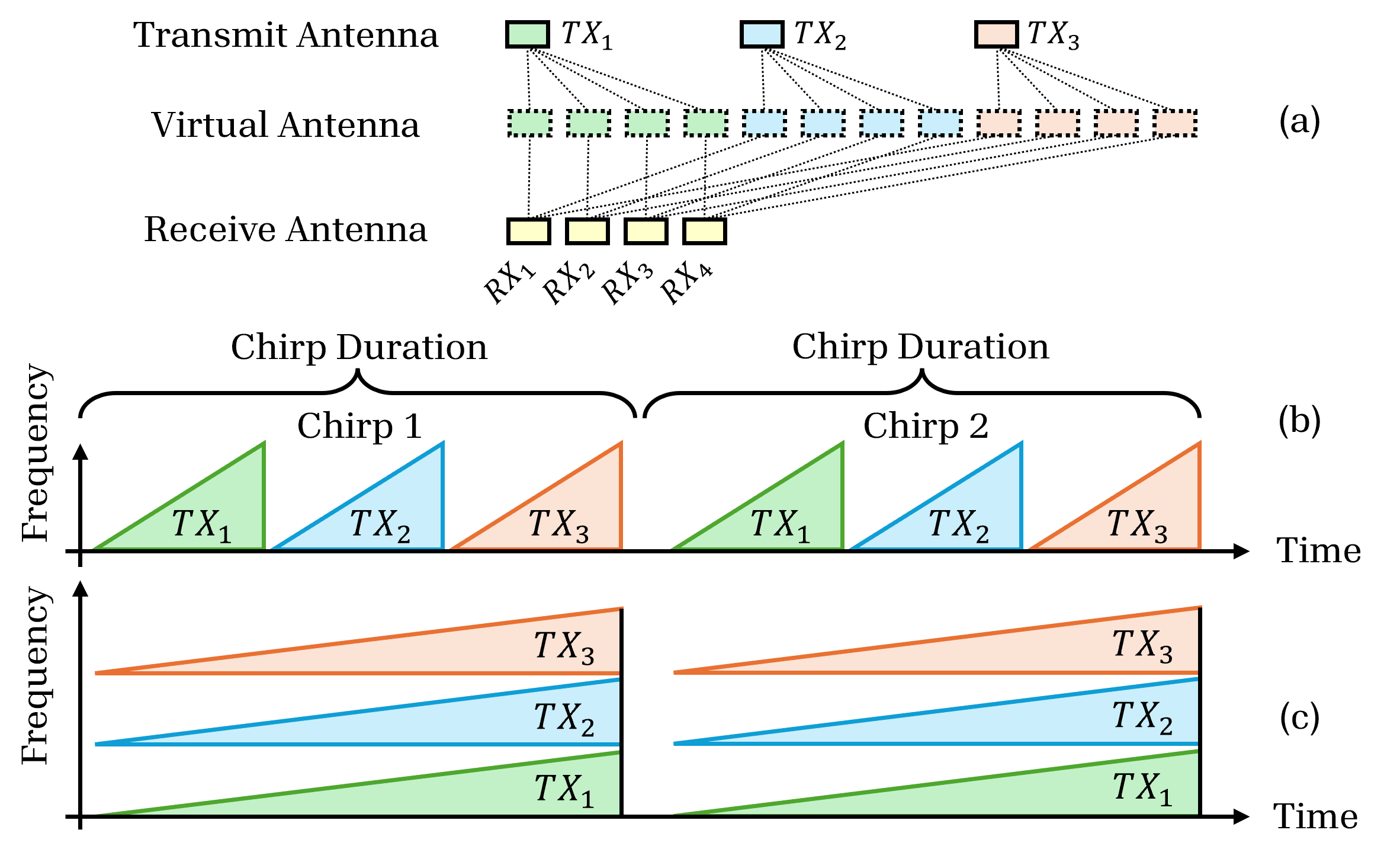}
    \caption{\textbf{MIMO radar virtual antenna formation and multiplexing.} (a) $N_{tx}$ transmitters and $N_{rx}$ receivers form $N_{tx}\!\times\!N_{rx}$ virtual antennas. RX channels read simultaneously. (b) TDM: TX elements fire sequentially. (c) DDM: TX elements fire spectrally interleaved FMCW pulses; virtual-array information is mixed in frequency per receiver.}
    \label{fig:MIMO_1}
\end{figure}

Sequential models that operate directly on streaming analog-to-digital converter (ADC) signals have emerged as a promising alternative: by processing chirps as they arrive, they reduce peak memory usage and can, in principle, make decisions earlier than frame-based models \cite{sharma2024chirpnet} (\Cref{fig:motivation} (a)). However, existing lightweight sequential approaches often struggle on more complex tasks such as object detection. We identify two key reasons. First, they typically compress or mix receiver channels early in the pipeline, discarding the explicit spatial localization information provided by a multiple-input multiple-output (MIMO) array. Second, in Doppler-division multiplexed (DDM) systems, they do not explicitly separate the contributions of different transmit antennas that are spectrally interleaved but remain latent in each receiver stream (\Cref{fig:MIMO_1}). Ignoring this structure causes the virtual-array elements of different transmitters to be aliased together, which degrades angle estimation and, in turn, detection accuracy.

We propose an efficient radar data processing architecture that keeps this MIMO structure explicit while remaining streaming-friendly (\Cref{fig:motivation} (b)). In a MIMO radar, $N_{tx}$ transmitter antennas and $N_{rx}$ receiver antennas form a large “virtual array”: each receiver element views a target with a distinct phase profile determined by the array geometry, and the combination of transmitter–receiver pairs yields many virtual elements with fine angular resolution (\Cref{fig:MIMO_1}). If the model mixes receiver channels too early or ignores which transmitter generated which echo, this virtual-array structure is lost, and recovering angle information becomes difficult, leading to degraded detection performance or the need for heavier decoders. To avoid this, we first process samples from each receiver channel independently so the encoder can learn per-antenna chirp features. We then introduce a lightweight cross-antenna attention module that learns how to combine per-chirp features across receiver channels using a small set of learnable virtual-array queries. This module effectively acts as a learnable beamformer: it reconstructs virtual-array features directly from the streaming signals without constructing range–angle–Doppler (RAD) tensors or using computationally expensive FFT based pipelines. The attention mixer adds negligible overhead to a streaming state space backbone while maintaining strong spatial localization information.

We also take advantage of the evolving nature of the scene motion between radar chirps to enable early decisions in object detection. Because adjacent chirps contribute primarily differential motion (Doppler) information, detection performance saturates after a small number of chirps in a radar frame, beyond which additional chirps yield diminishing returns. We therefore train with early-chirp supervision and deploy a calibrated stopping rule that triggers as soon as the latent state of the sequence model stabilizes, reducing both encoder FLOPs and latency.


The key contributions of this paper are:

\textbf{1. Physics-inspired spatial mixing for streaming ADC:} A cross-antenna attention module following independent RX processors that separates latent TX structure (DDM) to recover virtual-MIMO cues without reconstructing full RAD cubes.

\textbf{2. Hybrid design for efficiency:} Optimal placement of a cross-attention module between chirp-modeling channel-SSMs and frame-encoding chirp-SSM for the best compromise between computation and spatial resolving capacity.

\textbf{3. Sub-frame low-latency detection:} A chirp-wise SSM backbone that updates online and supports early decision via a calibrated stopping rule on chirp states, reducing encoder FLOPs and end-to-end latency (\Cref{fig:motivation} (c)).

\section{Related Work} \label{sec:related}

Classical object detection pipelines in radar vision first extract sparse point clouds (PC) via CFAR from range/angle/Doppler tensors and then run point-based or pseudo-image detectors (e.g. PointPillars variants) \cite{scheiner2021object,lang2019pointpillars}. While bandwidth-friendly and easy to fuse, these miss weak returns and often require multi-frame accumulation, which increases latency and can create motion ghosts. To preserve structural density, recent works process range--Doppler maps, range--angle heatmaps, or full RAD tensors with CNNs/Transformers \cite{dalbah2024transradar,rebut2022radial,giroux2023tfftradnet,palffy2020cnn}. However, these 3D grids (e.g., $256\!\times\!64\!\times\!12$ in a 3TX 4 RX configuration) are costly to transfer and infer on \cite{scheiner2021object}. For high-antenna imaging radars, forming and processing full RAD cubes in real time is especially demanding \cite{rebut2022radial}. These limitations motivate sequential ADC processing models that avoid constructing large tensors and reduce peak memory/latency.

End-to-end models on time-domain ADC aim to learn task-optimal transforms beyond fixed FFTs \cite{zhang2023adcnet}. Chirp-wise sequential encoders further reduce peak memory by updating an internal state as samples arrive; ChirpNet, for instance, reports $\sim\!15\times$ fewer parameters than CNN baselines \cite{sharma2024chirpnet}. However, many lightweight sequential designs compress or mix receiver (RX) channels early, weakening spatial localization and angle cues latent in the MIMO array. Deep state space models (SSMs) offer linear-time streaming with long-range dependence via structured state updates \cite{gu2021s4,smith2023simplified,gu2024mamba}, making them natural time-series backbones; yet without \emph{explicit} cross-antenna correlation, even SSM-based encoders risk discarding geometry that originates from the radar physics.

Beyond radar, there is a broad literature on \emph{anytime} and early-exit inference for deep models. CNNs such as MSDNet add intermediate classifiers with confidence- or entropy-based stopping to trade accuracy for compute on a per-input basis \cite{huang2018multiscale}. Transformer variants (DeeBERT, FastBERT) attach lightweight heads and use entropy/consistency criteria to decide when to stop \cite{xin-etal-2020-deebert,liu2020fastbert}. 
Radar-specific work has also begun to exploit temporal structure across frames for better perception \cite{li2022exploiting}, but typically assumes full-frame access and dense feature maps, whereas our focus is on chirp-wise, sub-frame decisions from streaming ADC.

In contrast, \textbf{RAVEN} combines these two: we design a radar physics inspired encoder that sequentially processes raw ADC that preserves MIMO structure via cross-antenna mixing and enables \emph{chirp-wise} early decisions by applying a calibrated stopping rule on the slow-time SSM state, rather than introducing heavy auxiliary heads. 




\section{Methodology} \label{sec:arch}

\subsection{Design Motivation}

We explicitly leverage the signal and array physics of FMCW MIMO radar when designing RAVEN’s encoder, instead of treating ADC samples as generic time-series data. In an FMCW radar, each target generates a beat frequency tied to its range and Doppler, and an $N_{\mathrm{rx}}$-element array encodes angle through deterministic phase shifts across antennas. These spatial phase patterns form the steering vector that enables angular resolution
\cite{singh2023mimofmcw, huang2011fmcw}.

Conventional sequential encoders often ignore this structure. For example, if each RX channel is reduced to a scalar and then averaged or passed through a shared $1\times1$ mix to form a per-chirp token, the operation is equivalent to applying a fixed uniform beamformer. This collapses the $N_{rx}$ dimensional receiver array response into a single value, discarding the relative phase differences that encode angle. Downstream layers then receive tokens stripped of spatial diversity, making angle recovery significantly harder.

The problem becomes worse in Doppler-division multiplexed (DDM) MIMO radars, where RX channels already contain linear mixtures of multiple TX waveforms. If the encoder does not explicitly preserve or disentangle these TX-specific components (akin to matched filters), early tokenization further mixes virtual-array responses. This additional entanglement degrades the network’s ability to learn angular structure and ultimately reduces detection accuracy.

This motivates two design choices in \textbf{RAVEN} :
\begin{itemize}
    \item \textbf{Per-RX fast-time processing:} maintain separate encoders for each RX channel so that per-antenna phase and amplitude structure is preserved.
    \item \textbf{Explicit cross-antenna mixing:} use a lightweight attention-based module that learns steering-like weights across RX channels and latent TX structure, instead of relying on implicit spatial learning in deep backbones.
\end{itemize}
These physics-driven constraints keep the encoder compact while retaining the spatial information needed for accurate localization.

\begin{figure*}[htb]
    \centering
    \includegraphics[width = \linewidth]{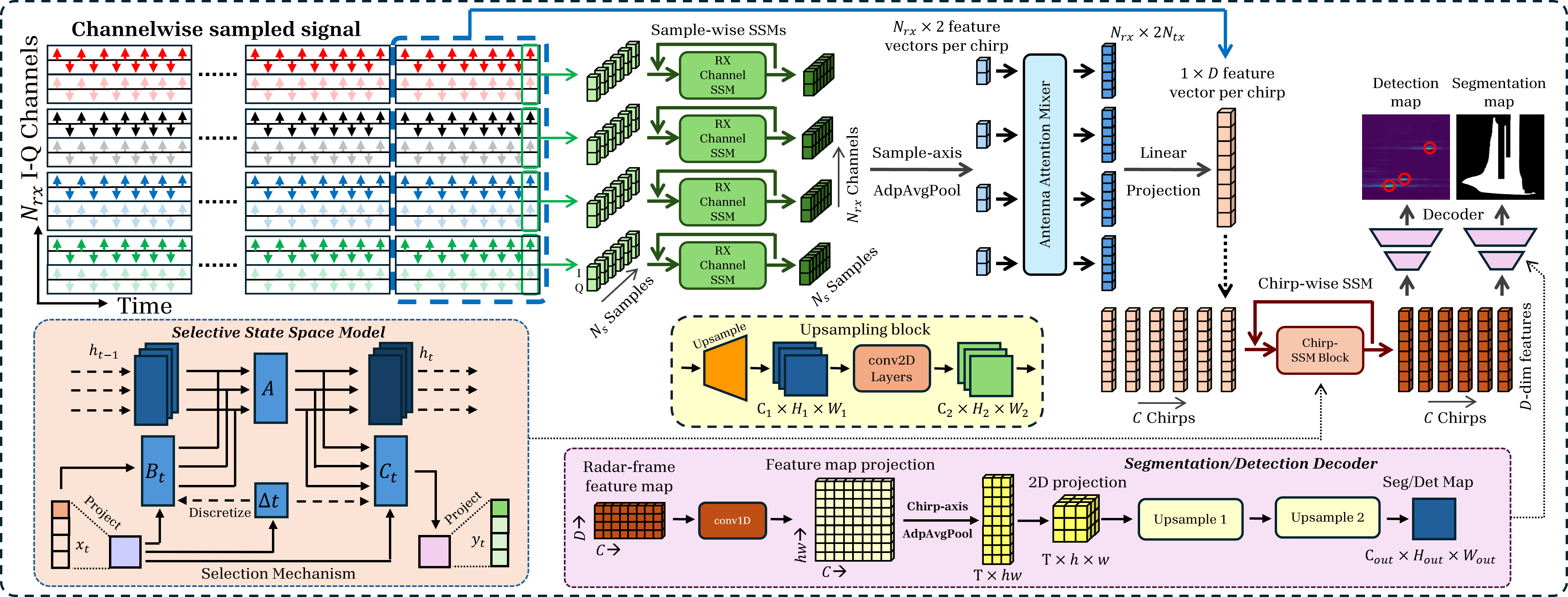}
    \caption{\textbf{RAVEN Architecture:} (1) Fast-time per-RX SSMs compress I/Q into compact 2-D tokens; (2) cross-antenna attention fuses RX channels and expands to virtual-MIMO features; (3) a chirp-wise SSM updates the state online across chirps; (4) a learned projection maps features to a $T\times H\times W$ grid; (5) lightweight decoders produce detection heatmaps/boxes and segmentation.}
    \label{fig:model_architecture}
\end{figure*}

\subsection{RAVEN Model Architecture}

RAVEN turns streaming ADC samples into BEV detections and freespace maps through five stages (\Cref{fig:model_architecture}):

\begin{itemize}
    \item \textbf{Fast-time SSMs:} each RX channel is processed independently by a small state space model to produce a compact token per RX and chirp.
    \item \textbf{Cross-antenna attention:} per-chirp RX tokens are fused with a lightweight attention module that learns spatial correlations and forms virtual MIMO features.
    \item \textbf{Chirp-wise SSM:} a slow-time SSM reads the chirp sequence, maintaining a hidden state so the model can update online and support anytime inference.
    \item \textbf{Spatial projection:} sequence features are mapped into a $T\times H\times W$ grid suitable for 2D decoding.
    \item \textbf{Lightweight decoders:} shallow CNN decoders produce detection heatmaps/boxes and freespace segmentation maps.
\end{itemize}
\vspace{-1em}
\paragraph{Notation.}
Each radar frame is a sequence of $N_c$ chirps (slow-time), and each chirp has $N_s$ fast-time samples. The receiver (RX) has $N_{\mathrm{rx}}$ channels and the transmitter (TX) has $N_{\mathrm{tx}}$ channels. We use complex (I/Q) samples, so the input channel dimension is $2N_{\mathrm{rx}}$. A frame is $$
\mathbf{X}\in\mathbb{R}^{N_c\times N_s\times 2N_{\mathrm{rx}}}
$$,
with axes (slow -time, fast-time, channels). Bold capitals denote tensors, bold lower-case vectors, and plain symbols denote dimensions. We write $\mathrm{LN}$ for LayerNorm, $\sigma(\cdot)$ for SiLU, and $\mathrm{Pool}_K$ for adaptive average pooling to length $K$.

\subsubsection{Parallel RX-channel SSM encoders (fast-time)}
For receiver $r\!\in\!\{1,\dots,N_{\mathrm{rx}}\}$ and chirp $k\!\in\!\{1,\dots,N_c\}$, let
\[
\mathbf{x}_{r,k}\in\mathbb{R}^{N_s\times 2}
\]
be the fast-time I/Q sequence for RX $r$ at chirp $k$. Each RX uses its own state space encoder $\mathrm{SSM}_r:\mathbb{R}^{N_s\times 2}\rightarrow\mathbb{R}^{N_s\times 2}$ (implemented with a Mamba block), and we compute
\[
\tilde{\mathbf{z}}_{r,k}=\mathrm{SSM}_r(\mathbf{x}_{r,k})\in\mathbb{R}^{N_s\times 2},
\mathbf{f}_{r,k}=\mathrm{Pool}_1\!\big(\tilde{\mathbf{z}}_{r,k}^{\top}\big)\in\mathbb{R}^{2}.
\]
Stacking all receivers gives
\[
\mathbf{F}_k=\big[\mathbf{f}_{1,k},\dots,\mathbf{f}_{N_{\mathrm{rx}},k}\big] \in \mathbb{R}^{N_{\mathrm{rx}}\times 2},
\mathbf{F}\in\mathbb{R}^{ N_c\times N_{\mathrm{rx}}\times 2}.
\]

\emph{Intuition.} Each RX stream is summarized to a tiny per-chirp token that still carries per-antenna range/phase information, providing a compact but geometry-aware input to the cross-antenna attention stage.

\subsubsection{Cross-antenna attention \& virtual MIMO expansion}
For chirp $k$, let $\mathbf{F}_k\!\in\!\mathbb{R}^{N_{\mathrm{rx}}\times 2}$ be the per-RX summaries from the fast-time SSMs. We first expand each RX to $d$-dim tokens and add a learnable RX embedding:
\[
\mathbf{H}^{\mathrm{rx}}_k
=\mathbf{W}_{\mathrm{in}}\mathbf{F}_k + \mathbf{E}^{\mathrm{rx}}
\;\in\;\mathbb{R}^{N_{\mathrm{rx}}\times d},
\mathbf{E}^{\mathrm{rx}}=[\mathbf{e}^{\mathrm{rx}}_{1},\ldots,\mathbf{e}^{\mathrm{rx}}_{N_{\mathrm{rx}}}]^{\!\top}.
\]

We introduce a bank of \emph{learnable TX queries}
$\mathbf{Q}\!\in\!\mathbb{R}^{N_{\mathrm{tx}}\times d}$ that probe the RX tokens via cross-attention
(queries $=$ TX, keys/values $=$ RX). With pre-norm,
\[
\mathbf{q}=\mathrm{LN}(\mathbf{Q}),\quad
\mathbf{k}=\mathrm{LN}(\mathbf{H}^{\mathrm{rx}}_k),\quad
\mathbf{v}=\mathbf{H}^{\mathrm{rx}}_k,
\]
the TX-updated tokens are
\[
\mathrm{Attn}(\mathbf{q},\mathbf{k},\mathbf{v})
=\mathrm{softmax}\!\left(\frac{\mathbf{q}\mathbf{k}^\top}{\sqrt{d}}\right)\mathbf{v}
\;\in\;\mathbb{R}^{N_{\mathrm{tx}}\times d}.
\]
We apply TX-side residual and feed-forward:
\[
\tilde{\mathbf{T}}
=\mathbf{Q}+\mathrm{Attn}(\mathbf{q},\mathbf{k},\mathbf{v})
\]
\[
\mathbf{T}=\tilde{\mathbf{T}}+\mathrm{FFN}\!\big(\mathrm{LN}(\tilde{\mathbf{T}})\big)
\;\in\;\mathbb{R}^{N_{\mathrm{tx}}\times d}.
\]

Next, for every $(r,t)$ pair we concatenate the corresponding RX and TX tokens and project to a compact two-dimensional feature:
\[
\mathbf{p}_{r,t}
=\mathbf{W}_{\mathrm{pair}}\,[\,\mathbf{h}^{\mathrm{rx}}_{r};\,\mathbf{t}_{t}\,]
\;\in\;\mathbb{R}^{2},\quad
\mathbf{W}_{\mathrm{pair}}\in\mathbb{R}^{2\times(2d)}.
\]
Stacking over $r$ and $t$ yields
$\mathbf{P}_k\in\mathbb{R}^{N_{\mathrm{rx}}\times N_{\mathrm{tx}}\times 2}$,
which we vectorize and normalize to form the per-chirp output:
\[
\mathbf{y}_k=\mathrm{LN}\!\big(\mathrm{vec}(\mathbf{P}_k)\big)
\;\in\;\mathbb{R}^{2N_{\mathrm{rx}}N_{\mathrm{tx}}}.
\]
Over all chirps this gives
\[
\mathbf{Y}=\big[\mathbf{y}_1,\ldots,\mathbf{y}_{N_c}\big]
\in\mathbb{R}^{ N_c\times (2N_{\mathrm{rx}}N_{\mathrm{tx}})}.
\]

\begin{figure}[htbp]
    \centering
    \includegraphics[width = \linewidth]{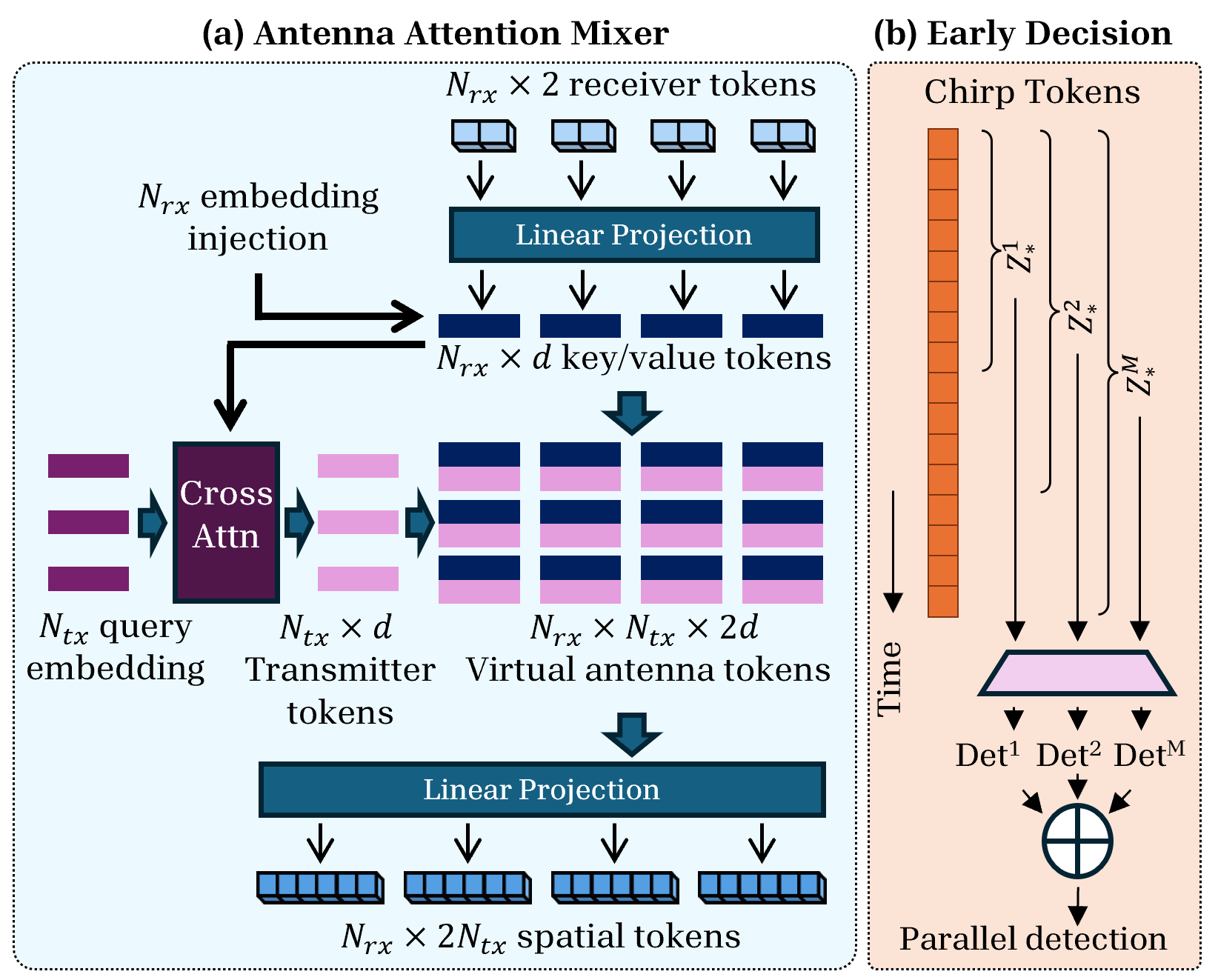}
    \caption{\textbf{(a) Attention Mixer:} Learnable transmitter queries are used to extract Doppler-division multiplexed information from the receiver signal in the time domain. These are fused together to form the virtual antenna array for retrieving the MIMO information. \textbf{(b) Early Decision Supervision:} During training, decoders take outputs from multiple chirp levels, and loss is computed simultaneously \cite{kusupati2022matryoshka}, forcing the model to converge on earlier chirps.}
    \label{fig:attention_mixer}
\end{figure}

\emph{Intuition.} In \Cref{fig:attention_mixer} (a) the TX queries act like learnable steering vectors that search the field of RX tokens, producing TX-specific summaries $\mathbf{T}$. Pairwise fusion $[\mathbf{h}^{\mathrm{rx}}_{r};\mathbf{t}_t]\mapsto\mathbb{R}^2$ then yields a compact feature for every virtual MIMO pair $(r,t)$, enabling the network to emphasize phase-consistent returns across antennas (DDM compatible) without constructing range–angle–Doppler (RAD) tensors.

\subsubsection{Chirp-wise (slow-time) SSM backbone}
We compress the channel dimension and prepare slow-time features:
\[
\mathbf{z}_k=\sigma\!\big(\mathbf{W}_{\mathrm{pre}}\;\sigma(\mathbf{W}_{\mathrm{red}}\mathbf{y}_k)\big)\in\mathbb{R}^{D},
\]
\[
\mathbf{Z}_\ast=[\mathbf{z}_1,\dots,\mathbf{z}_{N_c}]\in\mathbb{R}^{ N_c\times D},
\]
with $\mathbf{W}_{\mathrm{red}}\in\mathbb{R}^{D\times (2N_{\mathrm{rx}}N_{\mathrm{tx}})}$, $\mathbf{W}_{\mathrm{pre}}\in\mathbb{R}^{D\times D}$ and an optional shallow MLP.

We use Mamba-style structured SSMs that support both streaming updates and parallel training. The final slow-time representation is
\[
\mathbf{Z}_\ast = \text{SSM}(\mathbf{Z})\in\mathbb{R}^{ N_c\times D}.
\]
\emph{Intuition:} the SSM keeps a compact state while reading chirps in order, enabling online/anytime decisions without needing the full frame.

\subsubsection{Encoder--decoder projection and heads}

\paragraph{Detection branch.}
We project slow-time features to a compact spatio–temporal grid and decode object heatmaps and box offsets.
\begin{align*}
\underbrace{\mathbf{U}=\mathrm{Conv1D}_{C\!=\!D\to HW}\!\big(\mathbf{Z}_\ast^{\top}\big)}_{\mathbf{U}\in\mathbb{R}^{ HW\times N_c}}
\\
\xrightarrow{\ \mathrm{Pool}_{T_{\mathrm{det}}}\ }
\mathbf{U}_{\mathrm{det}}\in\mathbb{R}^{ HW\times T_{\mathrm{det}}}
\\
\xrightarrow{\ \mathrm{reshape}\ }
\mathbf{S}_{\mathrm{det}}\in\mathbb{R}^{ T_{\mathrm{det}}\times H\times W}.
\end{align*}
A shallow Conv–LN–SiLU stack with bilinear upsampling maps $\mathbf{S}_{\mathrm{det}}$ to a high-resolution feature map $\mathbf{A}$, from which $1{\times}1$ heads predict classification scores and box offsets, yielding $\mathrm{Det}=[\mathbf{P},\mathbf{R}]$:
\[
\mathbf{P}=\mathrm{sigmoid}\!\big(\mathrm{Conv}_{\text{cls}}(\mathbf{A})\big),\quad
\mathbf{R}=\mathrm{Conv}_{\text{reg}}(\mathbf{A}),
\]

\paragraph{Segmentation branch.}
We use an analogous projection to produce features for drivable-area (freespace) segmentation.
An analogous projection with temporal pooling of length $T_{\mathrm{seg}}$ produces $\mathbf{S}_{\mathrm{seg}}\in\mathbb{R}^{ T_{\mathrm{seg}}\times H\times W}$, followed by a similar Conv–LN–SiLU + upsampling stack to obtain segmentation logits
\[
\mathbf{M}=\mathrm{Conv}\big(\mathrm{Up}(\sigma(\mathrm{LN}(\mathrm{Conv}(\mathbf{S}_{\mathrm{seg}}))) )\big)
\]

\subsection{Sub-frame Decision Framework}
\label{sec:sub-frame-training}
For a constant-velocity target, information about $v$ accumulates via coherent computation across the chirps.
Doppler FFT needs $N_c$ chirps to achieve velocity resolution $\Delta v$, but detection often tolerates coarser $\Delta v$. This observation suggests that a sequential detector need not wait for all $N_c$ chirps: it can stop once its internal state has stabilized to a sufficient prediction.

\subsubsection{Training approach to enable sub-frame decision}

We implement this idea via multi-prefix supervision (\Cref{fig:attention_mixer}-b). Let $\mathcal{L}=\{L_1,\dots,L_M\}\subseteq\{1,\dots,N_c\}$ be a set of chirp-prefix lengths (with $L_M=N_c$).
For a frame, the encoder produces $\mathbf{Z}_\ast\in\mathbb{R}^{ N_c\times D}$.
For each $L\in\mathcal{L}$ we take the prefix $\mathbf{Z}_\ast^{(L)}=\mathbf{Z}_\ast[:,1{:}L,:]\in\mathbb{R}^{ L\times D}$ and pass it through the same projection and decoders to obtain
\[
\widehat{\mathrm{Det}}^{(L)} \in \mathbb{R}^{ 3\times H'\times W'},\qquad
\widehat{\mathrm{Seg}}^{(L)} \in \mathbb{R}^{ 1\times H''\times W''}.
\]
$\ell_{\text{det}}$ combines classification and box regression, and $\ell_{\text{seg}}$ is segmentation loss. All prefixes are supervised against the same ground-truth targets $(\mathrm{Det}^{\star},\mathrm{Seg}^{\star})$ for the frame, yielding a deep-supervision objective
\begin{align*}
\mathcal{L}_{\text{task}}
=\sum_{L\in\mathcal{L}}\Big[\,
\ell_{\text{det}}\big(\widehat{\mathrm{Det}}^{(L)},\,\mathrm{Det}^{\star}\big) \\
+\ell_{\text{seg}}\big(\widehat{\mathrm{Seg}}^{(L)},\,\mathrm{Seg}^{\star}\big)\,\Big]
\end{align*}



\subsubsection{Early inference rule}

Let $\mathbf{Z}_\ast^{(L)}=\{z_1,\dots,z_L\}\in\mathbb{R}^{L\times D}$ denote the chirp-wise latent states. For each new chirp $z_L$, we measure its novelty relative to the earlier bag of chirps via the minimum cosine distance
\[
d_L \;=\; \min_{1\le j < L}\!\Big(1 - \tfrac{z_L^\top z_j}{\|z_L\|\,\|z_j\|}\Big).
\]
When $d_L$ falls below a calibrated threshold $\tau$ (\Cref{fig:early_decision}), the latent dynamics have saturated and additional chirps offer negligible benefit. Because the decoder operates on blocks of $K$ pooled chirps, we compute a block-averaged score
\[
\bar d_m \;=\; \frac{1}{K}\sum_{L=(m-1)K+1}^{mK} d_L,
\]
and select the earliest block satisfying $\bar d_m \le \tau$. The final early-exit index is therefore
\[
L_{\mathrm{exit}} \;=\; K\,\min\{\,m:\,\bar d_m \le \tau\,\}.
\]

\section{Experimental Results} \label{sec:exp}


\subsection{Datasets} \label{sec:data}

\paragraph{RaDICaL (Radar, Depth, IMU, Camera).}
RaDICaL~\cite{lim2021radical} provides synchronized 77\,GHz frequency-modulated continuous-wave (FMCW) radar, stereo RGB-D, and inertial measurement unit (IMU) measurements. The radar uses a 4\,RX\,$\times$\,2\,TX TDM-MIMO configuration. For each frame we use complex ADC cube with $(N_c, N_s, N_{\mathrm{rx}}) = (64, 192, 8)$, where TX-RX pairs are interleaved along $N_{rx}$
The frame labels are generated from synchronized camera detections: tiled images are processed by RetinaNet~\cite{lin2017focalloss,chen2019fully}, and detections are merged into bird’s-eye view (BEV) occupancy masks.

\paragraph{RADIal (High-Definition Radar Multi-Task)}
RADIal~\cite{rebut2022radial} contains about two hours of synchronized driving with RGB video, a 16-beam LiDAR, and a 77\,GHz imaging radar over 91 sequences (urban, highway, rural). The radar uses a 12\,TX\,$\times$\,16\,RX DDM configuration (192 virtual antennas). Around 8{,}252 of $\sim$25{,}000 frames are annotated with vehicle centroids in polar and Cartesian coordinates and drivable-area (freespace) masks obtained from LiDAR. Since the TX are Doppler division multiplexed, the complex ADC cubes have size
$(N_c, N_s, N_{\mathrm{rx}}) = (256, 512, 16)$. We follow an 80/20 train/validation split during training process.




\subsection{Baselines and Implementation Details}

We compare \textbf{RAVEN} against radar-only CNN/FFT-based and UNet-style models~\cite{rebut2022radial,ronneberger2015unet,yang2018pixor,zhang2023adcnet}, attention/Transformer-based architectures including FFT–Transformer hybrids~\cite{giroux2023tfftradnet,dalbah2024transradar}, chirp-wise sequential models such as ChirpNet ~\cite{sharma2024chirpnet}, and ultra-light SSM-based encoders such as SSMRadNet~\cite{sen2025ssmradnetsamplewisestatespace}. All baselines are trained on the same ADC representation as RAVEN.

For \textbf{RADIal}, we train jointly for drivable-area segmentation and vehicle detection using Adam (learning rate $1\times10^{-4}$, weight decay $5\times10^{-6}$), batch size 8, and 200 epochs; segmentation uses a Jaccard (IoU) loss and detection uses Focal loss plus Smooth L1 regression. For \textbf{RaDICaL}, we train for BEV occupancy segmentation with Adam (learning rate $1\times10^{-4}$, weight decay $5\times10^{-6}$), batch size 8, and 300 epochs, using binary cross-entropy (BCE) as the primary loss.

\begin{figure*}[t]
    \centering
    \includegraphics[width=\linewidth]{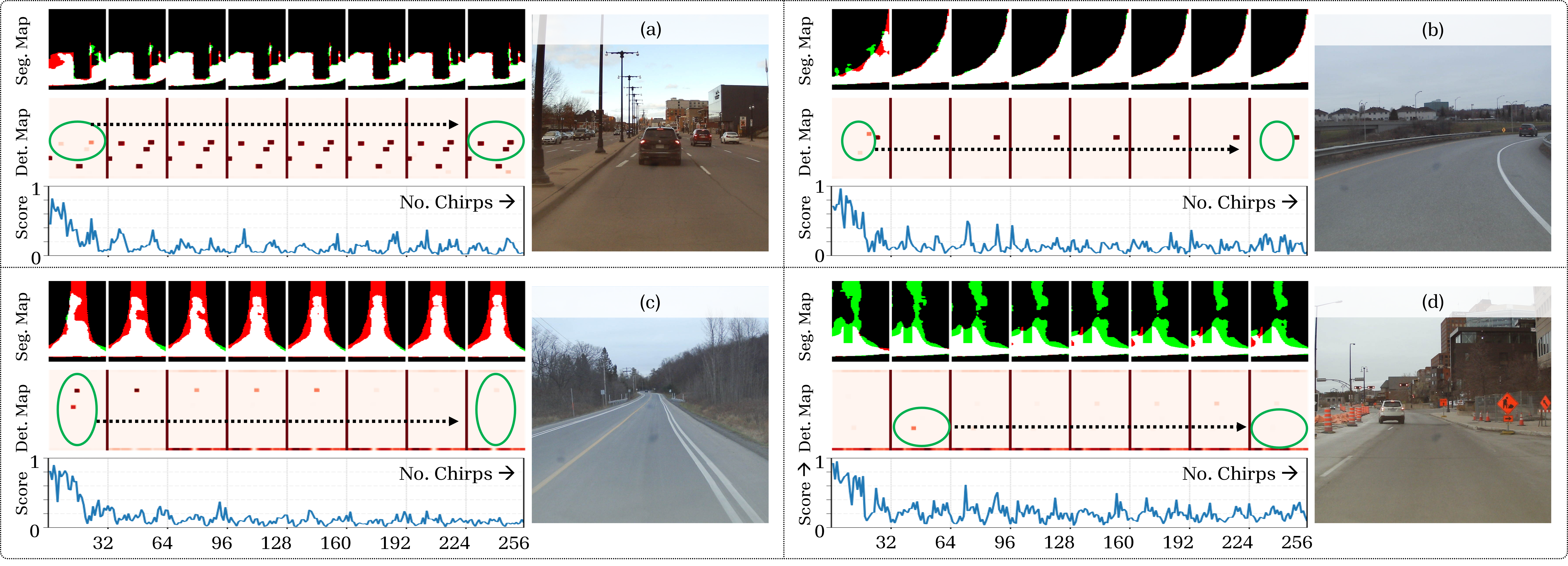}%

    \caption{Qualitative ablation of the adaptive decision module across four scenarios. Each example shows the RGB view, segmentation evolution over chirps (white - true positive, green - false positive, red - false negative), detection evolution (point-level RA predictions), and the chirp-state contribution signal. \textbf{Sample (a)} complex multi-vehicle scene where early-chirp assumptions about distant objects are refined into accurate detections. \textbf{Sample (b)} early-chirp false positives (“hallucinated” obstacles) are suppressed as more chirps arrive. \textbf{Sample (c)} early hallucinations fade but segmentation remains unreliable throughout. \textbf{Sample (d)} an object briefly emerges in clutter before vanishing, and a noisy chirp-similarity score depicts the irregularity of the data, resulting in poor segmentation and detection.}
    \label{fig:qualitative}
\end{figure*}

\begin{figure*}[htbp]
    \centering
    \includegraphics[clip, trim=0.5cm 1cm 2cm 1cm, height=5cm]{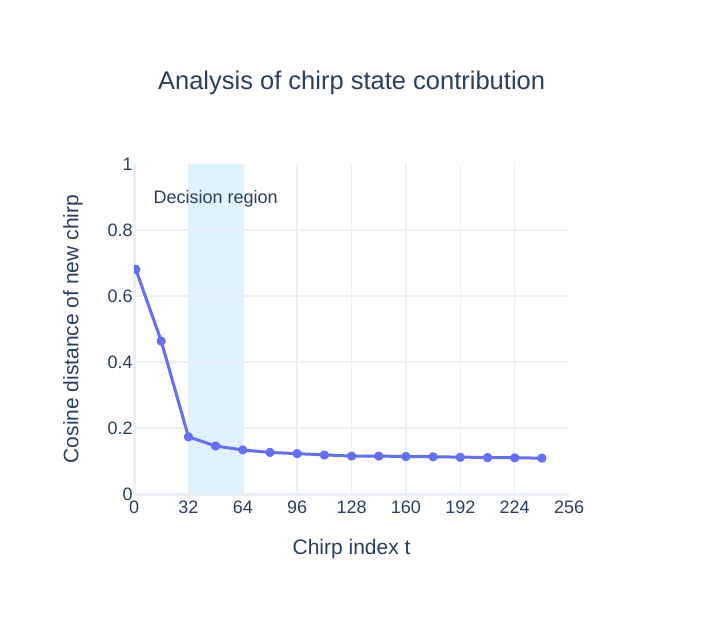}%
    \includegraphics[clip, trim=0.2cm 0.2cm 0.5cm 0.2cm, height=5.5cm]{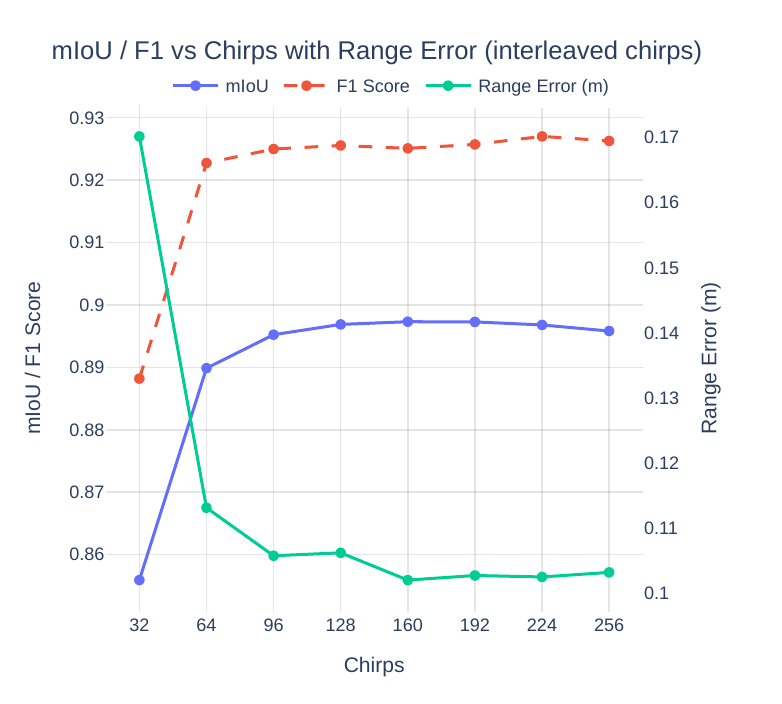}
    \includegraphics[clip, trim=0.8cm 1cm 1.5cm 1cm, height=5cm]{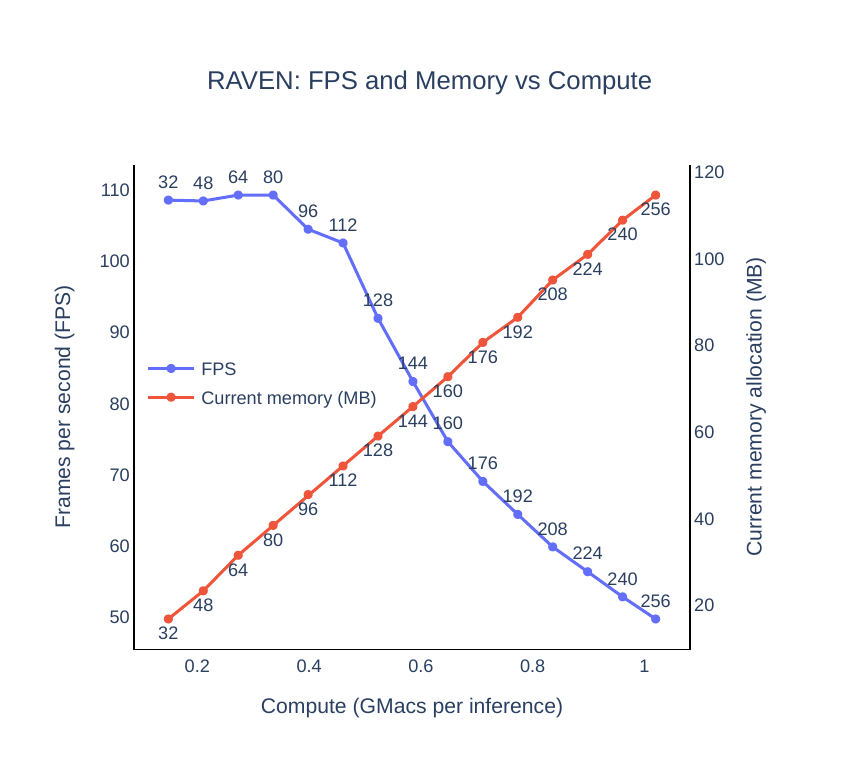}

    \caption{Design motivation for adaptive chirp selection. (Left) Minimum cosine-distance aggregate across all frames in train-set reveals a clear knee point beyond which new chirps add little novel information. (Middle) Validation-set scores show consistent gains from 32→64 chirps, with negligible improvements thereafter. (Right) Memory and latency scale strongly with chirp count, showing that reducing chirps provides substantial efficiency gains with minimal performance loss.}
    \label{fig:early_decision}
\end{figure*}

\subsection{Evaluation Metrics}


\quad\textbf{Segmentation:}
On RADIal, we follow~\cite{rebut2022radial} and report mean intersection-over-union (\textbf{mIoU}) for drivable-area masks. For RaDICaL we additionally report \textbf{Dice} coefficient measuring overlap between predicted and ground-truth masks and \textbf{Chamfer distance (CD) -} measuring the average bidirectional nearest-neighbor Euclidean distance between occupied pixels (or points) in the predicted and ground-truth BEV masks~\cite{fan2016point,yao2023radarcamreview}. While mIoU and Dice capture area overlap, 
Chamfer distance explicitly evaluates contour and boundary alignment. Using both Dice and Chamfer thus evaluates \emph{how much} of the area is correct and \emph{how well} object boundaries are localized.

\textbf{Detection:} 
For RADIal detection we report mean average precision (\textbf{mAP}), mean average recall (\textbf{mAR}), and \textbf{F1}-score following the official protocol~\cite{rebut2022radial}. 

\textbf{Efficiency:} Computational efficiency is summarized by multiply–accumulate operations (MACs), parameter count in millions (M), and end-to-end latency per frame measured on an NVIDIA RTX 4060 mobile GPU.

\subsection{Qualitative Results}

\Cref{fig:qualitative} illustrates how the adaptive decision module behaves across diverse scenarios. The chirp cosine distance score has a general downward trend in the first few chirps, after which it drops below a threshold value. In structured scenes with multiple vehicles, early chirps form coarse hypotheses that later chirps refine into stable detections, while inconsistent early hallucinations are naturally suppressed as more chirps accumulate. Conversely, cluttered or noisy scenes expose the limits of early-stage inference: segmentation may remain unreliable, objects may briefly appear and disappear, and the chirp-state contribution signal becomes erratic when the underlying data quality is poor. These examples highlight how the module integrates temporal evidence to stabilize predictions while avoiding unnecessary processing.

\Cref{fig:early_decision} motivates this design by showing that chirp-wise information exhibits diminishing returns. The feature cosine distance analysis across train-set (average cosine distance score) reveals a downward saturation trend with a distinct knee point where new chirps become highly redundant, enabling a natural threshold $\tau=0.2$ for early stopping. Consistent with this, performance trends on the validation set show clear improvements only up to roughly 64 chirps, after which mIoU and F1 gains flatten. At the same time, both memory usage and inference latency scale closely with chirp count—dropping the chirp budget from 256 to the 32--64 range yields more than a $2\times$ speedup with minimal accuracy loss. Together, these trends justify using an adaptive chirp-termination strategy to maintain accuracy while reducing computation.

\subsection{Quantitative Results}

\begin{table*}[t]
  \centering
  \scriptsize
  \setlength{\tabcolsep}{2.0pt}

  \begin{threeparttable}
    \caption{Comparison with prior works on \textbf{RaDICaL \cite{lim2021radical}}. 
    Best values per row are highlighted in \textbf{bold}.}
    \label{tab:radical_comparison}

    \begin{tabular}{@{}lccccccccc@{}}
      \toprule
      \textbf{Metric / Model} &
      \textbf{ChirpNet} &
      \textbf{ChirpNetLite} &
      \textbf{ChirpNet-SSM} &
      \textbf{ChirpNet-Attn} &
      \textbf{T-FFTRadNet} &
      \textbf{FFT-RadNet} &
      \textbf{UNet} &
      \textbf{SSMRadNet} &
      \textbf{\textcolor{blue}{RAVEN}} \\
       &
      \cite{sharma2024chirpnet} &
      \cite{sharma2024chirpnet} &
      \cite{chirpfamily} &
      \cite{chirpfamily} &
      \cite{giroux2023tfftradnet} &
      \cite{rebut2022radial} &
      \cite{ronneberger2015unet} &
      \cite{sen2025ssmradnetsamplewisestatespace} &
      \textcolor{blue}{(Ours)}
      \\
      \midrule

      \multicolumn{10}{c}{\textbf{Computational Complexity}} \\
      \midrule

      GMACs $\downarrow$ &
        1.480 &
        0.320 &
        0.340 &
        0.350 &
        15.990 &
        41.740 &
        15.140 &
        0.108 &
        \textcolor{blue}{\best{0.053}} \\
      Params (M) $\downarrow$ &
        3.780 &
        3.761 &
        3.761 &
        3.761 &
        9.000 &
        4.250 &
        17.270 &
        0.566 &
        \textcolor{blue}{\best{0.347}} \\
      \midrule

      \multicolumn{10}{c}{\textbf{Accuracy Metrics}} \\
      \midrule

      Dice Coefficient $\uparrow$ &
        0.986 &
        0.989 &
        0.990 &
        0.991 &
        0.995 &
        0.996 &
        0.996 &
        0.996 &
        \textcolor{blue}{\best{0.997}} \\
      Chamfer $\downarrow$ &
        0.097 &
        0.095 &
        0.088 &
        0.091 &
        0.108 &
        \best{0.076} &
        0.078 &
        0.086 &
        \textcolor{blue}{0.082} \\
      \bottomrule
    \end{tabular}

    \begin{tablenotes}[flushleft]
      \footnotesize
      \item Note:
      Dice coefficient: higher is better ($\uparrow$); Chamfer distance: lower is better ($\downarrow$).
    \end{tablenotes}
  \end{threeparttable}

  \setlength{\tabcolsep}{3.2pt}
\end{table*}

\begin{table*}[t]
\centering
\scriptsize
\begin{threeparttable}
\caption{Overall segmentation and detection performance on \textbf{RADIal\cite{rebut2022radial}}. Best values (global) in \textbf{bold}.}
\label{tab:results}
\begin{tabular}{l l c c c c c c c c c}
\toprule
\textbf{Class} 
  & \textbf{Model} 
  & \textbf{mIoU}\,$\uparrow$
  & \textbf{F1}\,$\uparrow$
  & \textbf{mAP}\,$\uparrow$
  & \textbf{mAR}\,$\uparrow$
  & \textbf{RE (m)}\,$\downarrow$
  & \textbf{AE ($^{\circ}$)}\,$\downarrow$
  & \textbf{GMACs}\,$\downarrow$
  & \textbf{Params (M)}\,$\downarrow$
  & \textbf{Lat. (ms)}\,$\downarrow$\\
\midrule
\multirow{13}{*}{Convolution} 
  & Pixor (PC) \cite{yang2018pixor}                & ---  & ---  & 0.96 & 0.32 & 0.17  & 0.25  & ---    & ---   & ---     \\
  & Pixor (RA) \cite{yang2018pixor}                & ---  & ---  & 0.96 & 0.82 & 0.12  & 0.20  & ---    & ---   & ---     \\
  & PolarNet \cite{nowruzi2020deep}                & 0.61 & ---  & ---  & ---  & ---   & ---    & ---    & ---   & ---     \\
  & Conv3D + FFT-RadNet \cite{wu2024sparseradnet}  & 0.75 & 0.47 & 0.58 & 0.39 & 0.19  & 0.33   & ---    & ---   & ---     \\
  & FFT-RadNet \cite{rebut2022radial}              & 0.74 & 0.88 & 0.97 & 0.82 & 0.14  & 0.17   & 146.82 & 3.80  & 53.59   \\
  & RLSM \cite{schubert2024radar}                  & 0.71 & 0.86 & 0.91 & 0.82 & ---   & ---    & ---    & ---   & ---     \\
  & FFT-RadUNet\tnote{a}                           & 0.75 & 0.80 & 0.83 & 0.77 & 0.16  & 0.10   & 134.40 & 18.48 & 44.92   \\
  & ADCNet \cite{zhang2023adcnet}                  & 0.79 & 0.89 & 0.93 & 0.86 & 0.14  & 0.11   & ---    & 2.50  & 18.13   \\
  & ADC\,UNet \cite{zhang2023adcnet}               & 0.77 & 0.85 & 0.88 & 0.82 & 0.18  & 0.11   & ---    & 17.50 & 8.18    \\
  & ADC\,UNet (NPT) \cite{zhang2023adcnet}         & 0.73 & 0.80 & 0.83 & 0.77 & 0.19  & 0.10   & ---    & ---   & ---     \\
  & FourierNet-FFT-RadUNet\tnote{b}                & 0.78 & 0.86 & 0.84 & 0.87 & 0.16  & 0.11   & 134.41 & 19.13 & 48.73   \\
  & FourierNet-FFT-RadNet\tnote{c}                 & 0.79 & 0.88 & 0.87 & 0.89 & 0.14  & 0.12   & 146.59 & 4.45  & 57.44   \\
  & C\text{--}M DNN \cite{cm-dnn}                  & 0.80 & 0.89  & 0.97 & 0.83 & 0.45  & ---    & 179.00 & 7.70  & 68.00   \\
\midrule
\multirow{4}{*}{Attention} 
  & ChirpNet (Self\text{-}Attn)\, \cite{chirpfamily}  & 0.65 & ---  & ---  & ---  & ---   & ---    & 33.00  & 50.95  & 20.37  \\
  & T-FFTRadNet \cite{giroux2023tfftradnet}        & 0.79 & 0.87 & 0.88 & 0.87 & 0.16  & 0.13   & 97.00  & 9.60  & 52.90   \\
  & TransRadar \cite{dalbah2024transradar}         & 0.82 & \textbf{0.93} & 0.95 & 0.91 & 0.15 & 0.10 & 171.50 & 3.70 & --- \\
  & EchoFusion \cite{liu2023echoes}                & ---  & 0.93  & \textbf{0.96} & \textbf{0.92} & 0.12  & 0.18   & ---    & ---   & ---     \\
\midrule
\multirow{1}{*}{Recurrent} 
  & ChirpNet (GRU) \cite{sharma2024chirpnet}       & 0.64 & ---  & ---  & ---  & ---   & ---    & 12.35  & 5.77  & 27.33   \\
\midrule
\multirow{2}{*}{SSM} 
  & ChirpNet (SSM)\, \cite{chirpfamily}    & 0.66 & ---  & ---  & ---  & ---   & ---    & 15.50  & 45.85  & 9.32    \\
  & SSMRadNet \cite{sen2025ssmradnetsamplewisestatespace} 
                                                    & 0.79 & 0.77 & 0.83 & 0.71 & 0.14  & 0.15   & 1.67 & \textbf{0.31} & 14.20   \\
\midrule
\multirow{1}{*}{GNN + Convolution} 
  & SparseRadNet \cite{wu2024sparseradnet}         & 0.78  & \textbf{0.93}  & \textbf{0.96} & 0.91 & 0.13  & 0.10   & 129.50 & 6.90  & ---   \\
\midrule
\multirow{2}{*}{SSM + Attention} 
  & {\color{blue}\textbf{RAVEN (Sub-frame, Ours)}}      
    & {\color{blue}0.85}
    & {\color{blue}0.89}
    & {\color{blue}0.88}
    & {\color{blue}0.89}
    & {\color{blue}0.17}
    & {\color{blue}0.25}
    & {\color{blue}\textbf{0.27}}
    & {\color{blue}1.51}
    & {\color{blue}\textbf{9.15}} \\
  & {\color{blue}\textbf{RAVEN (Full Frame, Ours)}}     
    & {\color{blue}\textbf{0.90}}
    & {\color{blue}\textbf{0.93}}
    & {\color{blue}0.95}
    & {\color{blue}\textbf{0.92}}
    & {\color{blue}\textbf{0.12}}
    & {\color{blue}\textbf{0.10}}
    & {\color{blue}\textbf{1.02}}
    & {\color{blue}1.51}
    & {\color{blue}20.08} \\
\bottomrule
\end{tabular}

\begin{tablenotes}\footnotesize
\item RE = range error; AE = azimuth error; RA = range–azimuth. $\uparrow$ higher is better; $\downarrow$ lower is better. Blue rows denote our models.
\item [a, b, c] [a] FFT-RadNet\cite{rebut2022radial}+UNet\cite{ronneberger2015unet}; FourierNet\cite{zhao2023cubelearn} FFT fed to [b] FFT-RadUNet, [c] FFT-RadNet.
\end{tablenotes}
\end{threeparttable}
\end{table*}

RAVEN delivers state-of-the-art (SOTA) performance on both \textbf{RADIal} and \textbf{RaDICaL} datasets. 

On RaDICaL~\cite{lim2021radical}, RAVEN achieves a Dice coefficient of \textbf{0.997} with Chamfer distance 0.082 at just 0.053 GMACs (see \Cref{tab:radical_comparison}). Compared to FFT\text{-}RadNet~\cite{rebut2022radial}, which reaches 0.996 Dice at 41.74 GMACs, this corresponds to nearly $790\times$ lower compute and about $12\times$ fewer parameters (0.35\,M vs.\ 4.25\,M), while maintaining near state-of-the-art mask quality and boundary alignment with a tiny fraction of the computational and parameter budget.

On RADIal~\cite{rebut2022radial}, our model attains an mIoU of \textbf{0.90}, F1 of \textbf{0.93}, and the lowest range/angle errors (RE 0.12\,m, AE $0.10^{\circ}$), while using only 1.02 GMACs (see \Cref{tab:results}), about $170\times$ less compute than TransRadar~\cite{dalbah2024transradar} (171.5 GMACs) and $95\times$ less than T\text{-}FFTRadNet~\cite{giroux2023tfftradnet} (97 GMACs), yet matching or surpassing their segmentation and detection accuracy, while enabling chirp-wise decisions at a fraction of the compute.





\section{Conclusion \& Future Work}\label{sec:conc}

We introduced RAVEN, an end-to-end machine learning architecture for radar-based perception that models FMCW radar chirp sequences using lightweight channel-wise and chirp-wise SSMs paired with cross-antenna attention. By aggregating information along the chirp dimension while preserving MIMO structure, RAVEN produces BEV freespace and object detections with state-of-the-art accuracy on RADIal and RaDICaL datasets, yet requires orders of magnitude fewer GMACs and parameters than prior radar-only models. Our analysis shows that the chirp-wise backbone yields stable prefix representations, causing the latent state to saturate early; this enables effective chirp subsampling and points toward compute-adaptive radar pipelines. Physics-aware sequential encoders can thus match heavy frame-based models under edge constraints. Looking forward, incorporating RAVEN into multimodal stacks and evaluating it across broader driving conditions offers a path toward robust, efficient multi-modal perception systems for real-world deployment. RAVEN therefore offers a practical, edge-friendly radar perception backbone for high-resolution object detection and freespace segmentation.
\section{Acknowledgement}\label{sec:ack}
This material is based upon work supported in part by SRC JUMP~2.0 (CogniSense, \#2023-JU-3133) and in part by DARPA through the OPTIMA program. Any opinions, or recommendations expressed in this material are those of the author(s) and do not reflect the views of SRC or DARPA. This research was also supported through research cyber infrastructure resources and services provided by the Partnership for an Advanced Computing Environment (PACE) at the Georgia Institute of Technology, Atlanta, Georgia, USA.

{
    \small
    \bibliographystyle{ieeenat_fullname}
    \bibliography{main}

@String(CVPR = {IEEE Conf. Comput. Vis. Pattern Recog.})

@String(ICCV = {Int. Conf. Comput. Vis.})

@ARTICLE{wang2021rodnet,
  author={Wang, Yizhou and Jiang, Zhongyu and Li, Yudong and Hwang, Jenq-Neng and Xing, Guanbin and Liu, Hui},
  journal={IEEE Journal of Selected Topics in Signal Processing}, 
  title={RODNet: A Real-Time Radar Object Detection Network Cross-Supervised by Camera-Radar Fused Object 3D Localization}, 
  year={2021},
  volume={15},
  number={4},
  pages={954-967},
  doi={10.1109/JSTSP.2021.3058895}}

@inproceedings{sharma2024chirpnet,
  title     = {ChirpNet: Noise-Resilient Sequential Chirp-Based Radar Processing for Object Detection},
  author    = {Sharma, Sudarshan and Kumawat, Hemant and Mukhopadhyay, Saibal},
  booktitle = {IEEE International Microwave Symposium},
  year      = {2024}
}

@InProceedings{giroux2023tfftradnet,
    author    = {Giroux, James and Bouchard, Martin and Laganiere, Robert},
    title     = {T-FFTRadNet: Object Detection with Swin Vision Transformers from Raw ADC Radar Signals},
    booktitle = {Proceedings of the IEEE/CVF International Conference on Computer Vision (ICCV) Workshops},
    month     = {October},
    year      = {2023},
    pages     = {4030-4039}
}

@InProceedings{rebut2022radial,
  author    = {Rebut, Julien and Ouaknine, Arthur and Malik, Waqas and P{\'e}rez, Patrick},
  title     = {Raw High-Definition Radar for Multi-Task Learning},
  booktitle = {Proceedings of the IEEE/CVF Conference on Computer Vision and Pattern Recognition (CVPR)},
  year      = {2022},
  pages     = {17000--17009},
  doi       = {10.1109/CVPR52688.2022.01651},
  note      = {Paper: \url{https://doi.org/10.1109/CVPR52688.2022.01651}. Dataset: \url{https://github.com/valeoai/RADIal}}
}

@article{wu2024sparseradnet,
  title   = {SparseRadNet: Sparse Perception Neural Network on Subsampled Radar Data},
  author  = {Wu, Jialong and Meuter, Mirko and Schöler, Markus and Rottmann, Matthias},
  journal = {arXiv preprint arXiv:2406.10600},
  year    = {2024}
}

@article{zhang2023adcnet,
  title   = {ADCNet: Learning from Raw Radar Data via Distillation},
  author  = {Zhang, Bo and Khatri, Ishan and Happold, Michael and Chen, Chulong},
  journal = {arXiv preprint arXiv:2303.11420},
  year    = {2023}
}

@misc{lim2021radical,
  title        = {RaDICaL: A Synchronized FMCW Radar, Depth, IMU and RGB Camera Dataset with Low-Level FMCW Radar Signals},
  author       = {Lim, Teck Yian and Markowitz, Spencer A. and Do, Minh N.},
  howpublished = {\url{https://doi.org/10.13012/B2IDB-3289560_V1}},
  year         = {2021}
}

@ARTICLE{zhao2023cubelearn,
  author={Zhao, Peijun and Lu, Chris Xiaoxuan and Wang, Bing and Trigoni, Niki and Markham, Andrew},
  journal={IEEE Internet of Things Journal}, 
  title={CubeLearn: End-to-End Learning for Human Motion Recognition From Raw mmWave Radar Signals}, 
  year={2023},
  volume={10},
  number={12},
  pages={10236-10249},
  doi={10.1109/JIOT.2023.3237494}}

@article{huang2011fmcw,
  author  = {Huang, Yanchuan and Brennan, Paul Victor and Patrick, Dave and Weller, I. and Roberts, Peters and Hughes, K.},
  title   = {{FMCW} Based {MIMO} Imaging Radar for Maritime Navigation},
  journal = {Progress In Electromagnetics Research},
  volume  = {115},
  pages   = {327--342},
  year    = {2011},
  doi     = {10.2528/PIER11021509},
  url     = {https://doi.org/10.2528/PIER11021509}
}

@inproceedings{nowruzi2020deep,
  title={Deep open space segmentation using automotive radar},
  author={Nowruzi, Farzan Erlik and Kolhatkar, Dhanvin and Kapoor, Prince and Al Hassanat, Fahed and Heravi, Elnaz Jahani and Laganiere, Robert and Rebut, Julien and Malik, Waqas},
  booktitle={2020 IEEE MTT-S International Conference on Microwaves for Intelligent Mobility (ICMIM)},
  pages={1--4},
  year={2020},
  organization={IEEE}
}

@InProceedings{sen2025ssmradnetsamplewisestatespace,
    author    = {Sen, Anuvab and Mohammad, Mir Sayeed and Mukhopadhyay, Saibal},
    title     = {SSMRadNet : A Sample-wise State-Space Framework for Efficient and Ultra-Light Radar Segmentation and Object Detection},
    booktitle = {Proceedings of the IEEE/CVF Winter Conference on Applications of Computer Vision (WACV)},
    month     = {March},
    year      = {2026},
    pages     = {4365-4374}
}

@INPROCEEDINGS{singh2023mimofmcw,
  author={Singh, Himali and Chattopadhyay, Arpan},
  booktitle={2023 31st European Signal Processing Conference (EUSIPCO)}, 
  title={Multi-Target Range and Angle Detection for MIMO-FMCW Radar with Limited Antennas}, 
  year={2023},
  volume={},
  number={},
  pages={725-729},
  doi=  {10.23919/EUSIPCO58844.2023.10289869}}

@ARTICLE{chirpfamily,
  author={Sharma, Sudarshan and Kumawat, Hemant and Sen, Anuvab and Park, Jinhyeok and Mukhopadhyay, Saibal},
  journal={IEEE Transactions on Radar Systems}, 
  title={Toward Efficient and Robust Sequential Chirp-Based Data-Driven Radar Processing for Object Detection}, 
  year={2025},
  volume={3},
  number={},
  pages={1435-1448},
  doi={10.1109/TRS.2025.3622514}}

@INPROCEEDINGS{schubert2024radar,
  author={Pushkareva, Mariia and Feldman, Yuri and Domokos, Csaba and Rambach, Kilian and Castro, Dotan Di},
  booktitle={2024 International Radar Conference (RADAR)}, 
  title={Radar Spectra-Language Model for Automotive Scene Parsing}, 
  year={2024},
  volume={},
  number={},
  pages={1-6},
  doi={10.1109/RADAR58436.2024.10993898}}

@article{kusupati2022matryoshka,
  title={Matryoshka representation learning},
  author={Kusupati, Aditya and Bhatt, Gantavya and Rege, Aniket and Wallingford, Matthew and Sinha, Aditya and Ramanujan, Vivek and Howard-Snyder, William and Chen, Kaifeng and Kakade, Sham and Jain, Prateek and others},
  journal={Advances in Neural Information Processing Systems},
  volume={35},
  pages={30233--30249},
  year={2022}
}

@inproceedings{liu2023echoes,
  title   = {Echoes Beyond Points: Unleashing the Power of Raw Radar Data in Multi-modality Fusion},
  author  = {Liu, Yang and Wang, Feng and Wang, Naiyan and Zhang, Zhaoxiang},
  booktitle = {Advances in Neural Information Processing Systems (NeurIPS)},
  year    = {2023}
}

@ARTICLE{cm-dnn,
  author={Jin, Yi and Deligiannis, Anastasios and Fuentes-Michel, Juan-Carlos and Vossiek, Martin},
  journal={IEEE Transactions on Intelligent Vehicles}, 
  title={Cross-Modal Supervision-Based Multitask Learning With Automotive Radar Raw Data}, 
  year={2023},
  volume={8},
  number={4},
  pages={3012-3025},
  doi={10.1109/TIV.2023.3234583}
}

@ARTICLE{lin2017focalloss,
  author={Lin, Tsung-Yi and Goyal, Priya and Girshick, Ross and He, Kaiming and Dollár, Piotr},
  journal={IEEE Transactions on Pattern Analysis and Machine Intelligence}, 
  title={Focal Loss for Dense Object Detection}, 
  year={2020},
  volume={42},
  number={2},
  pages={318-327},
  doi={10.1109/TPAMI.2018.2858826}}

@article{ronneberger2015unet,
  title   = {U-Net: Convolutional Networks for Biomedical Image Segmentation},
  author  = {Ronneberger, Olaf and Fischer, Philipp and Brox, Thomas},
  journal = {Medical Image Computing and Computer Assisted Intervention},
  year    = {2015},
  pages   = {234--241}
}

@inproceedings{gu2021s4,
  title     = {Efficiently Modeling Long Sequences with Structured State Spaces},
  author    = {Gu, Albert and Goel, Karan and R{\'e}, Christopher},
  booktitle = {International Conference on Learning Representations},
  year      = {2022},
  url       = {https://openreview.net/forum?id=uYLFoz1vlAC}
}

@inproceedings{yang2018pixor,
  title={Pixor: Real-time 3d object detection from point clouds},
  author={Yang, Bin and Luo, Wenjie and Urtasun, Raquel},
  booktitle={Proceedings of the IEEE conference on Computer Vision and Pattern Recognition},
  pages={7652--7660},
  year={2018}
}

@INPROCEEDINGS{chen2019fully,
  author={Tian, Zhi and Shen, Chunhua and Chen, Hao and He, Tong},
  booktitle={2019 IEEE/CVF International Conference on Computer Vision (ICCV)}, 
  title={FCOS: Fully Convolutional One-Stage Object Detection}, 
  year={2019},
  volume={},
  number={},
  pages={9626-9635},
  doi={10.1109/ICCV.2019.00972}}

@ARTICLE{yao2023radarcamreview,
  author={Yao, Shanliang and Guan, Runwei and Huang, Xiaoyu and Li, Zhuoxiao and Sha, Xiangyu and Yue, Yong and Lim, Eng Gee and Seo, Hyungjoon and Man, Ka Lok and Zhu, Xiaohui and Yue, Yutao},
  journal={IEEE Transactions on Intelligent Vehicles}, 
  title={Radar-Camera Fusion for Object Detection and Semantic Segmentation in Autonomous Driving: A Comprehensive Review}, 
  year={2024},
  volume={9},
  number={1},
  pages={2094-2128},
  doi={10.1109/TIV.2023.3307157}}

@INPROCEEDINGS{fan2016point,
  author={Fan, Haoqiang and Su, Hao and Guibas, Leonidas},
  booktitle={2017 IEEE Conference on Computer Vision and Pattern Recognition (CVPR)}, 
  title={A Point Set Generation Network for 3D Object Reconstruction from a Single Image}, 
  year={2017},
  volume={},
  number={},
  pages={2463-2471},
  doi={10.1109/CVPR.2017.264}}

@inproceedings{
huang2018multiscale,
title={Multi-Scale Dense Networks for Resource Efficient Image Classification},
author={Gao Huang and Danlu Chen and Tianhong Li and Felix Wu and Laurens van der Maaten and Kilian Weinberger},
booktitle={International Conference on Learning Representations},
year={2018},
url={https://openreview.net/forum?id=Hk2aImxAb},
}

@inproceedings{xin-etal-2020-deebert,
    title = "{D}ee{BERT}: Dynamic Early Exiting for Accelerating {BERT} Inference",
    author = "Xin, Ji  and
      Tang, Raphael  and
      Lee, Jaejun  and
      Yu, Yaoliang  and
      Lin, Jimmy",
    editor = "Jurafsky, Dan  and
      Chai, Joyce  and
      Schluter, Natalie  and
      Tetreault, Joel",
    booktitle = "Proceedings of the 58th Annual Meeting of the Association for Computational Linguistics",
    month = jul,
    year = "2020",
    address = "Online",
    publisher = "Association for Computational Linguistics",
    url = "https://aclanthology.org/2020.acl-main.204/",
    doi = "10.18653/v1/2020.acl-main.204",
    pages = "2246--2251"
}

@inproceedings{liu2020fastbert,
  title={Fastbert: a self-distilling bert with adaptive inference time},
  author={Liu, Weijie and Zhou, Peng and Wang, Zhiruo and Zhao, Zhe and Deng, Haotang and Ju, Qi},
  booktitle={Proceedings of the 58th annual meeting of the association for computational linguistics},
  pages={6035--6044},
  year={2020}
}

@article{han20234d,
  title={4d millimeter-wave radar in autonomous driving: A survey},
  author={Han, Zeyu and Wang, Jiahao and Xu, Zikun and Yang, Shuocheng and He, Lei and Xu, Shaobing and Wang, Jianqiang and Li, Keqiang},
  journal={arXiv preprint arXiv:2306.04242},
  year={2023}
}

@ARTICLE{fan20244d,
  author={Fan, Lili and Wang, Junhao and Chang, Yuanmeng and Li, Yuke and Wang, Yutong and Cao, Dongpu},
  journal={IEEE Transactions on Intelligent Vehicles}, 
  title={4D mmWave Radar for Autonomous Driving Perception: A Comprehensive Survey}, 
  year={2024},
  volume={9},
  number={4},
  pages={4606-4620},
  doi={10.1109/TIV.2024.3380244}
  }

@article{zhang2023perception,
  title={Perception and sensing for autonomous vehicles under adverse weather conditions: A survey},
  author={Zhang, Yuxiao and Carballo, Alexander and Yang, Hanting and Takeda, Kazuya},
  journal={ISPRS Journal of Photogrammetry and Remote Sensing},
  volume={196},
  pages={146--177},
  year={2023},
  publisher={Elsevier}
}

@inproceedings{li2022exploiting,
  title={Exploiting temporal relations on radar perception for autonomous driving},
  author={Li, Peizhao and Wang, Pu and Berntorp, Karl and Liu, Hongfu},
  booktitle={Proceedings of the IEEE/CVF Conference on Computer Vision and Pattern Recognition},
  pages={17071--17080},
  year={2022}
}

@article{paek2022k,
  title={K-radar: 4d radar object detection for autonomous driving in various weather conditions},
  author={Paek, Dong-Hee and Kong, Seung-Hyun and Wijaya, Kevin Tirta},
  journal={Advances in Neural Information Processing Systems},
  volume={35},
  pages={3819--3829},
  year={2022}
}

@ARTICLE{burnett2022we,
  author={Burnett, Keenan and Wu, Yuchen and Yoon, David J. and Schoellig, Angela P. and Barfoot, Timothy D.},
  journal={IEEE Robotics and Automation Letters}, 
  title={Are We Ready for Radar to Replace Lidar in All-Weather Mapping and Localization?}, 
  year={2022},
  volume={7},
  number={4},
  pages={10328-10335},
  doi={10.1109/LRA.2022.3192885}}

@article{patole2017automotive,
  title={Automotive radars: A review of signal processing techniques},
  author={Patole, Sujeet Milind and Torlak, Murat and Wang, Dan and Ali, Murtaza},
  journal={IEEE Signal Processing Magazine},
  volume={34},
  number={2},
  pages={22--35},
  year={2017},
  publisher={IEEE}
}

@article{sun2020mimo,
  title={MIMO radar for advanced driver-assistance systems and autonomous driving: Advantages and challenges},
  author={Sun, Shunqiao and Petropulu, Athina P and Poor, H Vincent},
  journal={IEEE Signal Processing Magazine},
  volume={37},
  number={4},
  pages={98--117},
  year={2020},
  publisher={IEEE}
}

@inproceedings{
smith2023simplified,
title={Simplified State Space Layers for Sequence Modeling},
author={Jimmy T.H. Smith and Andrew Warrington and Scott Linderman},
booktitle={The Eleventh International Conference on Learning Representations },
year={2023},
url={https://openreview.net/forum?id=Ai8Hw3AXqks}
}

@inproceedings{gu2024mamba,
  title={Mamba: Linear-time sequence modeling with selective state spaces},
  author={Gu, Albert and Dao, Tri},
  booktitle={First conference on language modeling},
  year={2024}
}

@article{scheiner2021object,
  author  = {Scheiner, Nicolas and Kraus, Florian and Appenrodt, Nils and Dickmann, J{\"u}rgen and Sick, Bernhard},
  title   = {Object Detection for Automotive Radar Point Clouds---A Comparison},
  journal = {AI Perspectives},
  volume  = {3},
  pages   = {6},
  year    = {2021},
  doi     = {10.1186/s42467-021-00012-z},
  url     = {https://doi.org/10.1186/s42467-021-00012-z}
}

@inproceedings{lang2019pointpillars,
  title={Pointpillars: Fast encoders for object detection from point clouds},
  author={Lang, Alex H and Vora, Sourabh and Caesar, Holger and Zhou, Lubing and Yang, Jiong and Beijbom, Oscar},
  booktitle={Proceedings of the IEEE/CVF conference on computer vision and pattern recognition},
  pages={12697--12705},
  year={2019}
}

@article{palffy2020cnn,
  title={CNN based road user detection using the 3D radar cube},
  author={Palffy, Andras and Dong, Jiaao and Kooij, Julian FP and Gavrila, Dariu M},
  journal={IEEE Robotics and Automation Letters},
  volume={5},
  number={2},
  pages={1263--1270},
  year={2020},
  publisher={IEEE}
}

@INPROCEEDINGS{tiling,
  author={Kumawat, Hemant and Mukhopadhyay, Saibal},
  booktitle={2022 International Joint Conference on Neural Networks (IJCNN)}, 
  title={Radar Guided Dynamic Visual Attention for Resource-Efficient RGB Object Detection}, 
  year={2022},
  volume={},
  number={},
  pages={1-8},
  doi={10.1109/IJCNN55064.2022.9892184}}

@InProceedings{dalbah2024transradar,
    author    = {Dalbah, Yahia and Lahoud, Jean and Cholakkal, Hisham},
    title     = {TransRadar: Adaptive-Directional Transformer for Real-Time Multi-View Radar Semantic Segmentation},
    booktitle = {Proceedings of the IEEE/CVF Winter Conference on Applications of Computer Vision (WACV)},
    month     = {January},
    year      = {2024},
    pages     = {353-362}
}
}
\newpage
\clearpage
\setcounter{page}{1}
\maketitlesupplementary
\begin{figure*}[t]
    \centering
    \includegraphics[width=\linewidth]{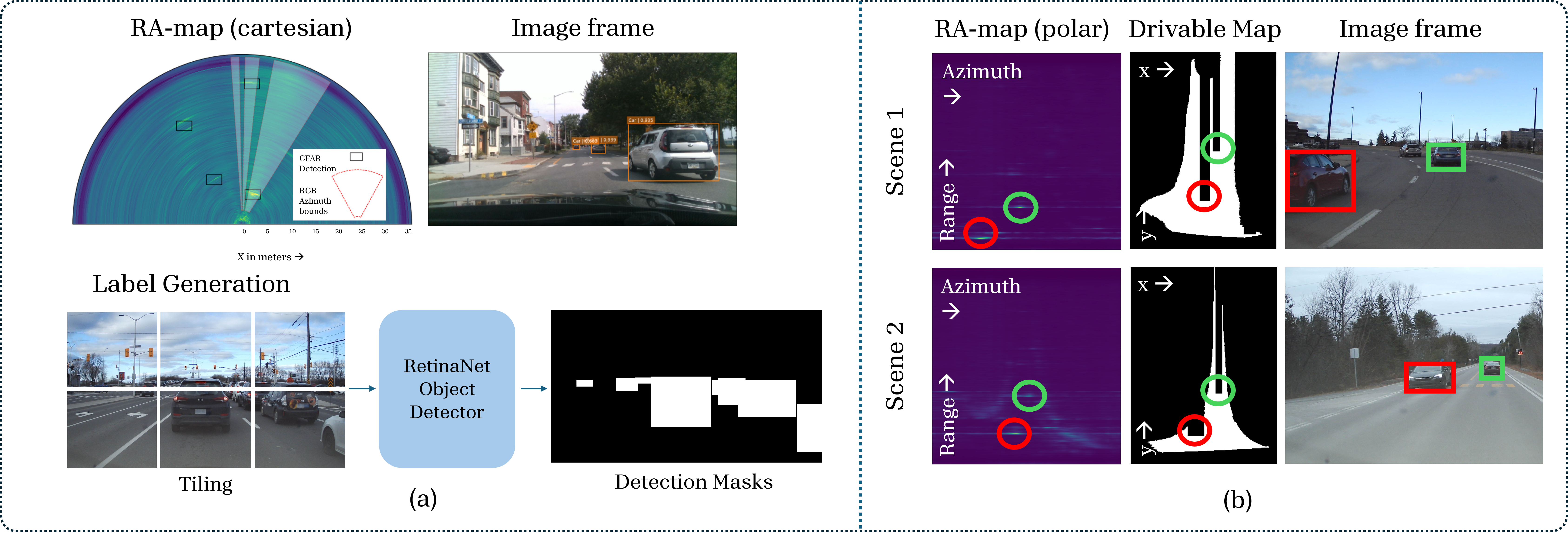}
    \caption{(a) \textbf{RaDICaL~\cite{lim2021radical}}: label generation from RGB frames using a tiled RetinaNet detector (adapted from~\cite{sharma2024chirpnet}). (b) \textbf{RADIal~\cite{rebut2022radial}}: FFT of raw ADC data produces range–azimuth maps; CFAR yields radar point clouds; segmentation maps mark drivable (white) vs.\ non-drivable (black) areas; nearest and second-nearest vehicles are highlighted in red and green, respectively.}
    \label{fig:dataset}
\end{figure*}

\section{Experimental Details}
\label{sec:additional_experiments}

\subsection{Datasets}
\label{subsec:dataset_description}

\subsubsection{RaDICaL dataset and annotation}
\label{subsubsec:radical}

We use the RaDICaL dataset~\cite{lim2021radical}, which provides synchronized measurements from a $4$-Rx, $3$-Tx $77$\,GHz FMCW radar, an RGB camera, a depth camera, and an inertial measurement unit (IMU). The depth camera produces reliable depth estimates only up to approximately $10$\,m, making it less effective for distant objects, whereas the radar remains sensitive to far-range targets. Scenes are recorded from a vehicle-mounted sensor rig across urban streets, country roads, and highways.

Unlike many prior radar datasets~\cite{yao2023radarcamreview}, RaDICaL releases raw ADC samples in addition to preprocessed range–Doppler or range–angle maps. This preserves the full semantic content of the radar data and enables efficient raw chirp-wise processing. While the radar hardware supports a $3$-Tx$\times$4-Rx MIMO configuration, the dataset was collected using a $2$-Tx$\times$4-Rx TDM MIMO setup, yielding $8$ virtual channels per chirp. Our pipeline uses all available virtual channels.          

\paragraph{RaDICaL Annotation pipeline.}
Supervised radar learning is limited by the difficulty of generating high-quality labels directly in the radar domain (e.g., range–azimuth–Doppler tensors or sparse point clouds). Instead of annotating radar data manually or relying on CFAR-based heuristics, we derive supervision from synchronized RGB images. We use a RetinaNet~\cite{lin2017focalloss} detector with a ResNet-50 backbone pre-trained on COCO on the camera images. To improve detection of small and distant objects, we adopt a tiling strategy~\cite{tiling}: each image is split into overlapping tiles, inference is run independently on each tile, and detections are stitched back into the original resolution. This improves recall for far objects compared to a single-pass detector. We restrict COCO classes to \emph{person}, \emph{bicycle}, \emph{car}, \emph{motorcycle}, \emph{bus}, and \emph{truck}.

From the stitched detections, we generate a binary mask in image space. This mask serves as the ground-truth signal during training. Importantly, we do \emph{not} use radar-to-camera calibration matrices to project annotations across modalities; instead, the model learns cross-modal alignment implicitly through its architecture. This avoids dependence on calibration, eliminates the need to store radar-domain labels, and minimizes the alignment noise and sparsity issues seen in RODNet-style labels~\cite{wang2021rodnet}. The tiling strategy can produce duplicate detections when a single object spans multiple tiles; we mitigate this with non-maximum suppression (NMS) after stitching. An overview of the RaDICaL annotation pipeline is shown in Fig.~\ref{fig:dataset}(a).

\subsubsection{RADIal dataset overview}
\label{subsubsec:radial}

For comparison and context, we briefly summarize the RADIal dataset~\cite{rebut2022radial}. RADIal uses a high-definition imaging radar with $N_{\text{Rx}} = 16$ receiver antennas and $N_{\text{Tx}} = 12$ transmitter antennas, giving $N_{\text{Rx}}N_{\text{Tx}} = 192$ virtual channels. This dense virtual array provides fine azimuth resolution and supports elevation estimation.

The radar is accompanied by a 16-layer automotive-grade LiDAR, a 5\,Mpix RGB camera mounted behind the windshield, and synchronized GPS and CAN traces for vehicle pose and kinematics. The three sensors have parallel horizontal lines of sight in the driving direction, and their extrinsic calibration is provided. RADIal contains $91$ sequences of $1$–$4$\,minutes each (city, highway, and countryside driving), for a total of roughly $25$k synchronized frames, of which $8{,}252$ frames are labeled with about $9{,}550$ vehicles. The RADIal signal-processing and labeling pipeline is summarized in Fig.~\ref{fig:dataset}(b).

\begin{figure*}[htbp]
    \centering
    \includegraphics[width=\textwidth]{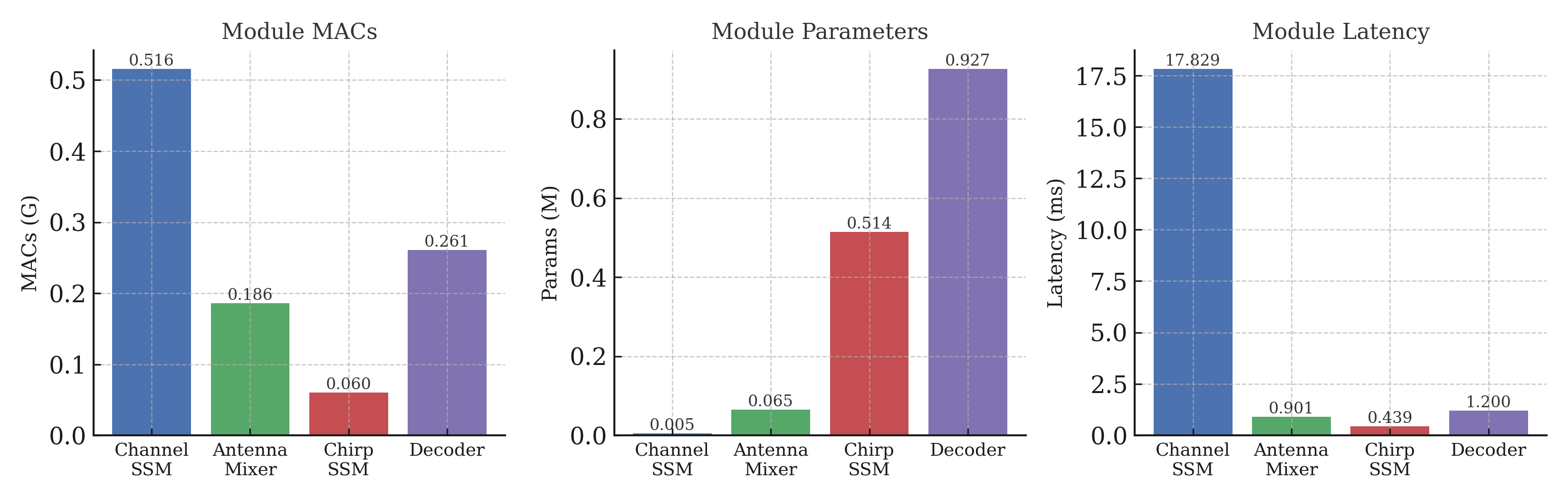}
    \caption{Per-block latency (ms) on a single GPU. The channel SSM is the main sequential bottleneck because it processes long fast-time sequences; the mixer and decoders are highly parallelizable.}
    \label{fig:block_latency}
\end{figure*}

Nevertheless, RADIal is the only large-scale dataset that provides raw analog-to-digital converter (ADC) radar signals rather than only preprocessed FFT cubes. This makes it possible to train foundation models directly on raw radar data streams, yielding competitive performance and architectures like RAVEN that can exploit raw signals efficiently for on-edge deployment.

\section{RAVEN Block-Wise Analysis}
\label{sec:blockwise}

RAVEN’s encoder–decoder pipeline consists of four logical components: (i) per-RX channel SSMs that operate along fast time, (ii) an antenna attention mixer that reconstructs virtual-MIMO features, (iii) a chirp-wise SSM backbone along slow time, and (iv) lightweight decoders for detection and segmentation. We profile them individually. Figure~\ref{fig:block_latency} summarizes the per-block parameter count, GMACs, and latency contributions, normalized to the full model. These plots show a consistent picture: most parameters reside in the 2D decoders, most MACs in the combination of chirp-wise SSM and decoders, and most latency in the channel SSM. The antenna mixer, despite encoding detailed virtual-MIMO structure, contributes only a small fraction of total compute and parameter count. This means we can afford 
spatial reasoning without compromising efficiency, provided that the fast-time block remains narrow and the slow-time backbone operates on sufficiently compressed tokens.
\vspace{-1em}



\section{Physics-guided Encoder Design}
\label{sec:physics}

The design of RAVEN’s encoder is guided directly by the signal and array physics
of FMCW MIMO radar. In this section, we move from the basic chirp model to the
virtual-array view and then to architectural choices: (i) how fast-time
structure suggests 1D state space models, (ii) how MIMO geometry encodes angle,
(iii) why naive channel mixing destroys that information, and (iv) how our
channel SSMs and antenna mixer modules implement a physics-aligned, end-to-end encoder for object detection.
Section~\ref{sec:ablation_channel_mixer_full} then validates these choices
empirically.        

\subsection{FMCW chirp and beat signal}

A single FMCW chirp of duration $T_c$ and bandwidth $B$ has instantaneous
transmit frequency
\begin{equation}
    f_{\text{tx}}(t) = f_0 + S t,
    \qquad S = \frac{B}{T_c},
    \quad 0 \le t \le T_c,
\end{equation}
and complex baseband signal
\begin{equation}
    s_{\text{tx}}(t)
    = \exp\!\left( j 2\pi \Big( f_0 t + \tfrac{1}{2} S t^2 \Big) \right).
\end{equation}
A target at range $R$ and radial velocity $v$ yields an echo delayed by
$\tau = \tfrac{2R}{c}$ and Doppler-shifted by $f_D = \tfrac{2v}{\lambda}$,
where $c$ is the speed of light and $\lambda$ is the carrier wavelength. After
mixing with the transmit signal and low-pass filtering, the resulting beat
signal can be approximated as
\begin{equation}
    s_b(t) \approx
    \exp\!\left( j 2\pi \big( f_b t + \text{const} \big) \right),
    \qquad
    f_b \approx f_r + f_D,
\end{equation}
with range-dependent frequency $f_r = \tfrac{2 S R}{c}$ and Doppler frequency
$f_D = \tfrac{2v}{\lambda}$. Thus, \emph{range} manifests as a linear frequency
along fast time, while \emph{velocity} appears as phase evolution across
chirps.

Let $T_s$ denote the ADC sampling period and
$n \in \{0,\dots,N_s{-}1\}$ the fast-time index. For chirp index $k$ with
repetition interval $T_R$, a single-target beat sample at one receiver is
approximately
\begin{equation}
    x_k[n]
    \propto \exp\!\big( j 2\pi (f_r n T_s + f_D k T_R) \big),
\end{equation}
which is the starting point for our fast-time state space encoders: the
fast-time dimension is a 1D sequence whose frequency encodes range, motivating
SSMs along fast time (ADC samples).

\subsection{MIMO virtual array and angle encoding}

For an $N_{\text{Rx}}$-element receive array with inter-element spacing $d$, a
plane wave from azimuth $\theta$ induces a spatial steering vector
\begin{equation}
    \mathbf{a}(\theta)
    =
    \big[\,1,
          e^{j\phi},
          \dots,
          e^{j (N_{\text{Rx}}-1)\phi}\big]^\top,
    \quad
    \phi = 2\pi \frac{d}{\lambda}\sin\theta.
\end{equation}
Stacking the beat samples across antennas for chirp $k$ and fast-time index $n$
yields
\begin{equation}
    \mathbf{x}_k[n]
    = \sum_{\ell} A_\ell
        e^{j 2\pi (f_{r,\ell} n T_s + f_{D,\ell} k T_R)}
        \mathbf{a}(\theta_\ell)
    + \mathbf{w}_k[n],
\end{equation}
where $A_\ell$ and $\theta_\ell$ denote the complex amplitude and angle of the
$\ell$-th target and $\mathbf{w}_k[n]$ represents noise and clutter.

In TDM/DDM MIMO, each receiver additionally sees echoes from multiple
transmitters, so the virtual array combines TX and RX patterns. The virtual
steering vector becomes a Kronecker product of TX and RX steering vectors, and
different transmitters are separated in either time (TDM) or Doppler (DDM).
Crucially, \emph{angle information is encoded in relative phase differences
across antennas and transmitters}; any operation that averages these channels
too early risks collapsing the array response to a single beam. Architecturally
this means we should preserve per-antenna channels until we have a mechanism
that can explicitly reason over them.

\subsection{Why naive channel mixing loses angle}

To see how early mixing harms angular resolution, consider a simplistic encoder
that first maps each receiver's fast-time sequence into a scalar summary and
then averages across receivers. Let $x_{r,k}[\cdot]$ denote the fast-time
samples for receiver $r$ and chirp $k$, and let $g(\cdot)$ be a (near-linear)
temporal encoder. We define
\begin{equation}
    u_{r,k} = g\!\big( x_{r,k}[\cdot] \big),
    \qquad
    \mathbf{u}_k
    =
    \big[ u_{1,k},\dots,u_{N_{\text{Rx}},k} \big]^\top,
\end{equation}
and obtain a per-chirp token via uniform averaging
\begin{equation}
    z_k
    = \tfrac{1}{N_{\text{Rx}}} \sum_{r=1}^{N_{\text{Rx}}} u_{r,k}
    = \mathbf{w}^\mathrm{H} \mathbf{u}_k,
    \qquad
    \mathbf{w} = \tfrac{1}{N_{\text{Rx}}} \mathbf{1}.
\end{equation}

If the scene is dominated by a single far-field target, then $\mathbf{u}_k$ is
approximately proportional to the steering vector $\mathbf{a}(\theta)$, so the
token becomes
\begin{equation}
    z_k \propto \mathbf{w}^\mathrm{H}\mathbf{a}(\theta)
    = \tfrac{1}{N_{\text{Rx}}} \mathbf{1}^\mathrm{H}\mathbf{a}(\theta).
\end{equation}
This is precisely the output of a fixed beamformer with weights $\mathbf{w}$:
all spatial information is compressed into one scalar, and only that one beam
pattern is available to the downstream network. Relative phase shifts
$e^{j r \phi}$ across antennas, which distinguish different angles $\theta$, no
longer appear explicitly in the representation.

In DDM/TDM MIMO, where TX waveforms are interleaved in Doppler or time, this
problem becomes more severe: the virtual array structure is already entangled
across chirps and frequencies, and early channel mixing further entangles it, making it difficult for later layers to recover angle-of-arrival (AoA) cues
without reconstructing RAD tensors. This motivates an encoder that \emph{first} models each channel's fast-time dynamics and \emph{then} performs explicit, learned spatial mixing across antennas.

\subsection{Per-RX Channel Fast Time SSMs and Antenna mixer as a radar physics-friendly alternative}

RAVEN avoids this pitfall by inserting two carefully structured stages before
the slow-time backbone.

\paragraph{Per-RX channel fast time SSMs:}
Instead of aggregating channels immediately, we maintain a separate fast-time
encoder for each receiver. For receiver $r$ and chirp $k$, we collect the I/Q
sequence
\begin{equation}
    \mathbf{x}_{r,k} \in \mathbb{R}^{N_s \times 2},
\end{equation}
and feed it to a Mamba-style state space model $\text{SSM}_r$:
\begin{align}
    \tilde{\mathbf{z}}_{r,k}
        &= \text{SSM}_r(\mathbf{x}_{r,k}) \in \mathbb{R}^{N_s \times 2}, \\
    \mathbf{f}_{r,k}
        &= \text{Pool}_1\!\big(\tilde{\mathbf{z}}_{r,k}^\top\big)
           \in \mathbb{R}^{2},
\end{align}
where $\text{Pool}_1$ adaptively averages the fast-time dimension to length~1.
Stacking across receivers yields
\begin{equation}
    \mathbf{F}_k
    = [\mathbf{f}_{1,k},\dots,\mathbf{f}_{N_{\text{Rx}},k}]
    \in \mathbb{R}^{N_{\text{Rx}}\times 2},
\end{equation}
so each antenna contributes a compact per-chirp descriptor that retains its
relative phase and amplitude structure. This implements the ``first compress
fast time per channel'' step suggested by the physics above.

\paragraph{Attention-based antenna mixer:}
The antenna mixer then interprets $\mathbf{F}_k$ as a set of tokens and learns
how to combine them in a way analogous to a set of learnable beams. After
projecting from $\mathbb{R}^2$ to $\mathbb{R}^d$ and adding RX embeddings, we
obtain
\begin{equation}
    \mathbf{H}_k^{\text{rx}} = W_{\text{in}} \mathbf{F}_k + \mathbf{E}^{\text{rx}}
    \in \mathbb{R}^{N_{\text{Rx}}\times d},
\end{equation}
and introduce $N_{\text{Tx}}$ TX queries
$\mathbf{Q} \in \mathbb{R}^{N_{\text{Tx}}\times d}$. Multi-head attention
produces a set of TX-aligned features
\begin{equation}
    \mathbf{T}_k
    = \text{Attn}(\mathbf{Q},\mathbf{H}_k^{\text{rx}},\mathbf{H}_k^{\text{rx}})
    \in \mathbb{R}^{N_{\text{Tx}}\times d},
\end{equation}
which can be interpreted as learnable steering patterns over the RX tokens.

To expose joint TX--RX information to the downstream SSM, we form small
pairwise features for every $(r,t)$ combination (e.g., by concatenation and a
linear layer) and compress each pair to a two-dimensional vector:
\begin{align}
    \mathbf{p}_{r,t,k}
        &= W_{\text{pair}}
           \big[\mathbf{h}^{\text{rx}}_{r,k};\mathbf{t}_{t,k}\big]
           \in \mathbb{R}^{2}, \\
    \mathbf{y}_k
        &= \text{LN}\!\left(\text{vec}\big(\{\mathbf{p}_{r,t,k}\}_{r,t}\big)\right)
           \in \mathbb{R}^{2 N_{\text{Rx}} N_{\text{Tx}}}.
\end{align}
Thus, each chirp is represented by a $2N_{\text{Rx}}N_{\text{Tx}}$-dimensional
\emph{learned virtual-antenna feature vector}. Crucially, this representation is
obtained directly from time-domain ADC signals through learned projections and
attention; we never perform explicit 2D/3D FFTs across range or angle, and we
never construct dense range--azimuth--Doppler (RAD) tensors.

Stacking over chirps gives
\begin{equation}
    \mathbf{Y}
    = [\mathbf{y}_1,\dots,\mathbf{y}_{N_c}]
    \in \mathbb{R}^{N_c \times (2 N_{\text{Rx}} N_{\text{Tx}})},
\end{equation}
which preserves the structure of the MIMO array in a compact latent space and
feeds directly into the chirp-wise SSM. This gives a physics-inspired
end-to-end encoder: fast-time SSMs for range, attention for angle, and
chirp-wise SSMs for temporal evolution.

\begin{table*}[t]
    \centering
    \small
    \begin{tabular}{lcccccc}
        \toprule
        Model Variant & Channel SSM & Antenna Mixer & mIoU & F1 & mAP & GMACs \\
        \midrule
        (A) Shared Fast-time SSM
            & \xmark & \xmark & 0.79 & 0.77 & 0.81 & 1.67 \\
        (B) Cross-Antenna Attention + Shared Fast-time SSM
            & \xmark & \xmark & 0.80 & 0.79 & 0.83 & 38.89 \\
        (C) Shared Fast-time SSM + Cross-Antenna Attention
            & \xmark & \cmark & 0.79 & 0.80 & 0.83 & 1.62 \\
        (D) Cross-Antenna Attention + Channel SSM
            & \cmark & \cmark & 0.84 & 0.88 & 0.88 & 34.56 \\
        \rowcolor{cvprblue!8}
        (E) \textcolor{cvprblue}{\textbf{Channel SSM + Cross-Antenna Attention (RAVEN, full-frame)}}
            & \textcolor{cvprblue}{\cmark}
            & \textcolor{cvprblue}{\cmark}
            & \textcolor{cvprblue}{\textbf{0.90}}
            & \textcolor{cvprblue}{\textbf{0.93}}
            & \textcolor{cvprblue}{\textbf{0.95}}
            & \textcolor{cvprblue}{\textbf{1.02}} \\
        \rowcolor{cvprblue!8}
        (F) \textcolor{cvprblue}{\textbf{Channel SSM + Cross-Antenna Attention (RAVEN, sub-frame)}}
            & \textcolor{cvprblue}{\cmark}
            & \textcolor{cvprblue}{\cmark}
            & \textcolor{cvprblue}{\textbf{0.85}}
            & \textcolor{cvprblue}{\textbf{0.89}}
            & \textcolor{cvprblue}{\textbf{0.88}}
            & \textcolor{cvprblue}{\textbf{0.27}} \\
        \bottomrule
    \end{tabular}  
    \caption{\textbf{Ablation of channel SSM and antenna mixer on RADIal.}
    All variants share the same chirp-wise SSM backbone and decoders.
    Our physics-guided RAVEN encoders (blue rows), which apply per-RX channel SSMs before the antenna mixer, achieve the best trade-off between accuracy
    and compute; the sub-frame variant further improves efficiency for
    early-exit decisions.}
    \label{tab:ablation_channel_mixer}
\end{table*}

\section{Ablation: Role and Ordering of Per RX Channel Fast Time SSM and Antenna Mixer}
\label{sec:ablation}

The radar physics discussion suggests that both the per-RX channel SSMs and the cross-antenna attention mixer are important, and that their ordering should
follow the natural flow of information. Our hypothesis is to first compress ADC samples across each receiver channel
along fast time, then isolate angle information from the channels. To validate this, we
compare the model variants in Table~\ref{tab:ablation_channel_mixer}, which all
share the same chirp-wise SSM and decoders but differ in how they model fast
time and cross-antenna interactions.

\begin{figure*}[t]
    \centering
    \includegraphics[width=\linewidth]{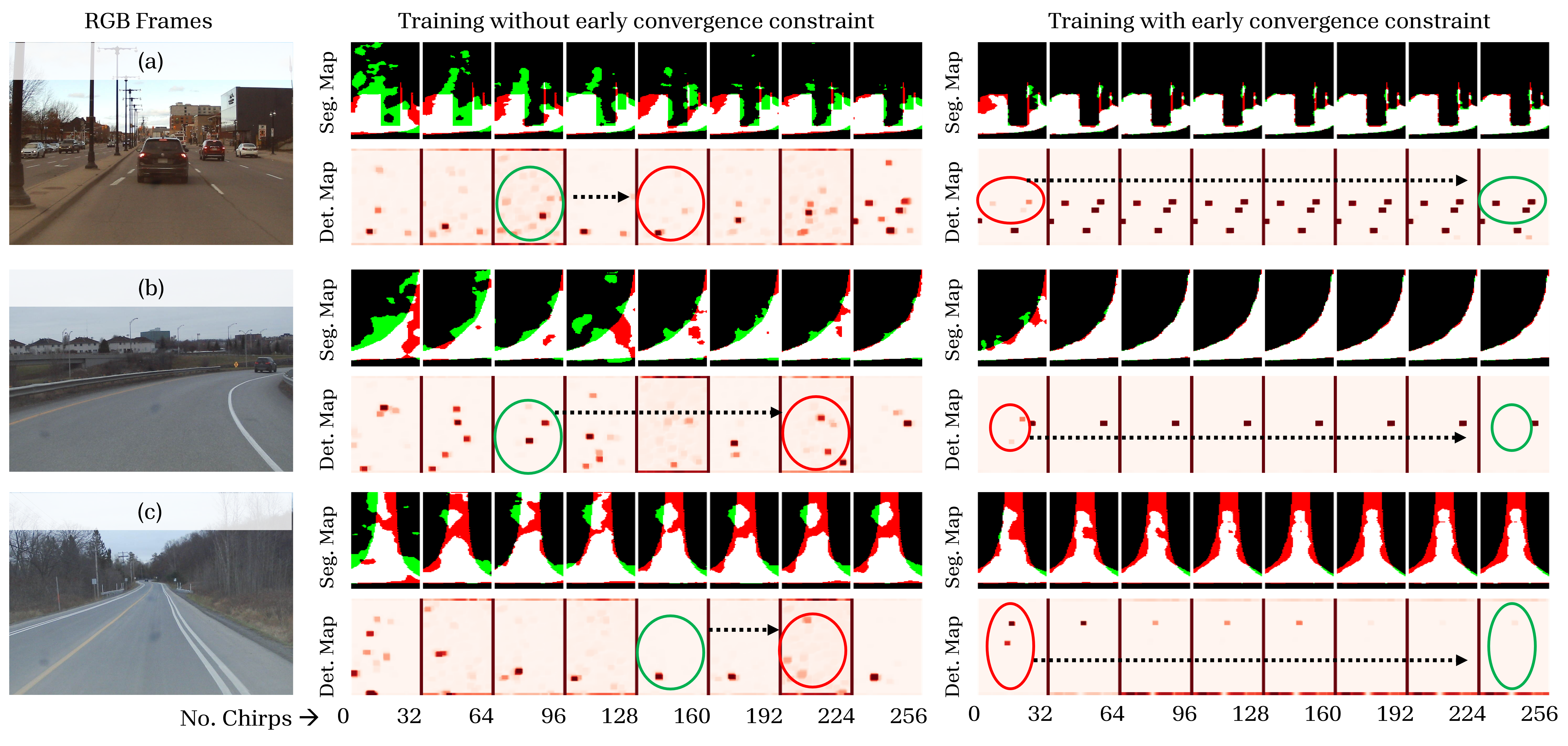}

    \caption{Segmentation and detection maps across driving scenes with and without multi-chirp supervision.
Without supervision across chirp levels, segmentation gradually approaches the ground truth, but detection remains unstable throughout the sequence, consistent with \Cref{fig:no_early_decision}. In (a), the model forgets real objects mid-frame that only reappear at the end. In (b), it initially identifies one set of objects but later predicts an entirely different set. In (c), it begins to hallucinate obstacles near the final chirps. With multi-chirp supervision, these issues disappear: detection becomes consistent across the sequence, and both segmentation and detection remain accurate through the final frame. }
    \label{fig:differences_in_states}
\end{figure*}

\paragraph{ Model Variants:}
We briefly restate what each row does and why some variants are much heavier:

\begin{itemize}
    \item \textbf{(A) Shared fast-time SSM.} A single fast-time SSM operates on
    all $2N_{\text{Rx}}$ input channels jointly. There is no per-RX channel SSM
    and no dedicated antenna mixer; the model treats the ADC samples as a
    generic multichannel sequence. This gives a reasonable baseline in both
    accuracy and compute (1.67 GMACs).

    \item \textbf{(B) Cross-antenna attention + shared fast-time SSM.} Here we
    augment the shared fast-time SSM with global cross-antenna interactions
    inside the same block, but still without a separate channel SSM module or a
    structured antenna mixer head. In our implementation, this attention is
    applied at \emph{full fast-time resolution}: it sees roughly
    $512 \times N_{\text{Rx}}$ tokens per chirp instead of a compressed set of
    per-antenna summaries. Because attention scales at least quadratically with
    the sequence length, this makes the block extremely expensive (38.89
    GMACs) even though the accuracy gain over (A) is small.

    \item \textbf{(C) Fast-time SSM + cross-antenna attention.} A shared
    fast-time SSM is followed by a dedicated cross-antenna attention mixer.
    This introduces an explicit mixer module, but because the fast-time SSM is
    still shared across all channels, it does not produce clean per-antenna
    summaries. The mixer therefore operates on features that partially blur
    channel structure, and accuracy improves only marginally over (A)/(B),
    while compute (1.62 GMACs) remains comparable to (A).

    \item \textbf{(D) Cross-antenna attention + channel SSM.} Both channel SSMs
    and the mixer are present, but in the \emph{reverse} order: cross-antenna attention is applied first on lightly projected I/Q samples, and
    the resulting mixed features are then processed by per-RX SSMs. As in (B),
    the attention still runs on long fast-time sequences ($\approx 512$
    samples per antenna), so it sees a large number of tokens and dominates the
    compute. This explains why (D) achieves good accuracy (mIoU~0.84, F1~0.88)
    but remains very heavy at 34.56 GMACs.

    \item \textbf{(E) Channel SSM + cross-antenna attention (RAVEN, full-frame).}
    Our proposed hybrid encoder: per-RX channel SSMs first compress each
    fast-time sequence into a low-dimensional token, so each chirp is
    represented by only $N_{\text{Rx}}$ channel tokens instead of
    $512 \times N_{\text{Rx}}$ time samples. The cross-antenna attention mixer then operates on this compressed set of tokens, reconstructing virtual MIMO
    structure at a much smaller sequence length. This ordering is motivated by the physics analysis in Section~\ref{sec:physics}. 

    \item \textbf{(F) Channel SSM + cross-antenna attention (RAVEN, sub-frame).}
    This variant uses the hybrid encoder as (E) but trains with our early-chirp criterion and decodes from a sub-frame subset of chirps. It maintains strong accuracy while reducing compute to 0.27 GMACs, providing an efficient early-exit option for on-edge deployment.
\end{itemize}

\subsection*{Optimal placement of the hybrid design for efficiency.}
The trends in Table~\ref{tab:ablation_channel_mixer} support the physics-guided
design. Variants (B) and (D) show that simply adding cross-antenna attention on
top of raw fast-time sequences is not a good trade-off: attention over
$\sim 512 \times N_{\text{Rx}}$ tokens per chirp is powerful but incurs tens of
GMACs. Variants (A) and (C), which avoid that extreme cost, either lack
structured cross-antenna reasoning or do not preserve clean per-channel
summaries and therefore underperform in accuracy. Our RAVEN encoders in (E)
and (F) implement a hybrid placement: channel SSMs first compress each
fast-time stream into a single token per antenna, and the cross-antenna attention mixer then operates on this compact set of tokens. This
reduces the attention sequence length by roughly a factor of 512 while
preserving the virtual-array structure, and is exactly why (E) and (F) achieve
the best balance between accuracy and compute, turning the SSM-first mixer
placement into a core physics-inspired architectural contribution rather than just another
variant.

\section{Early Chirp State Saturation Experiment}
\label{sec:ablation_channel_mixer_full}
\begin{figure}[htbp]
    \centering

    \begin{subfigure}[t]{\linewidth}
        \centering
        \includegraphics[clip, trim=0.2cm 0.2cm 0.5cm 0.2cm, height=5.5cm]{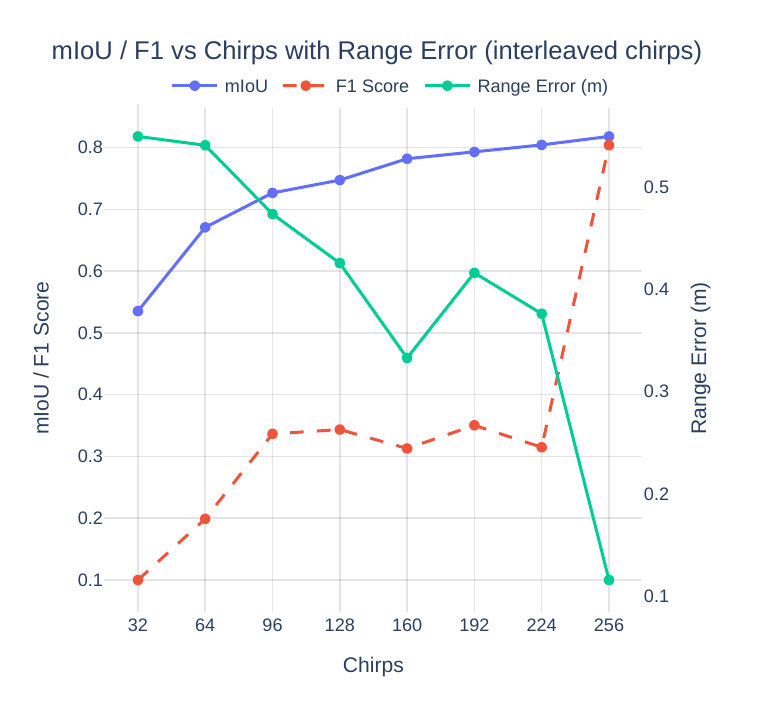}
        \caption{}
        \label{fig:no_early_decision:a}
    \end{subfigure}

    \vspace{0.3cm}

    \begin{subfigure}[t]{\linewidth}
        \centering
        \includegraphics[clip, trim=0.2cm 0.2cm 0.5cm 0.2cm, height=5.5cm]{images/ScoresVSchirps.pdf}
        \caption{}
        \label{fig:no_early_decision:b}
    \end{subfigure}

    \caption{Design motivation for adaptive chirp selection.
\textbf{(a)} Validation curves illustrate how detection and segmentation performance evolve with increasing chirp count. Without explicit early-convergence constraints, detection remains suboptimal until the full chirp frame is processed.
\textbf{(b)} Introducing multi-chirp supervision during training encourages the model to saturate earlier and learn temporal continuity across chirps, yielding smoother convergence and higher overall detection–segmentation performance.}
    \label{fig:no_early_decision}
\end{figure}

We evaluate the impact of enforcing early state convergence by decoding from partial chirp sets, compared to training without this constraint. Consistent with the observations in \cite{sen2025ssmradnetsamplewisestatespace}, we find that mIoU improves rapidly in the early chirp regime. However, this behavior does not naturally extend to detection, where performance depends on information accumulated across the full chirp sequence (\Cref{fig:no_early_decision:a}). To encourage the latent states to saturate earlier—enabling reliable early-exit decisions—we decode intermediate outputs from multiple chirp subsets and supervise each with detection targets (Section 3.3):
\begin{align*}
\mathcal{L}_{\text{task}}
=\sum_{L\in\mathcal{L}}\Big[\,
\ell_{\text{det}}\big(\widehat{\mathrm{Det}}^{(L)},\,\mathrm{Det}^{\star}\big) \\
+\ell_{\text{seg}}\big(\widehat{\mathrm{Seg}}^{(L)},\,\mathrm{Seg}^{\star}\big)\,\Big]
\end{align*}
Supervising detection at multiple chirp depths forces the model to extract complete spatial cues from early temporal observations. Learning this temporal–spatial continuity within radar frames improves overall performance, as shown in \Cref{fig:no_early_decision:b}.

%
%
We further visualize the effect of this training strategy on sub-frame decisions (\Cref{fig:differences_in_states}). Without early-decision supervision, the temporal progression of the radar frame is poorly preserved: detection heatmaps fluctuate significantly across chirps, with the model sometimes forgetting strong reflectors or hallucinating obstacles mid-frame. With the early-chirp constraint, these inconsistencies largely disappear. Both detection and segmentation exhibit smoother evolution across chirps, and the latent states stabilize much earlier. This leads not only to improved detection performance but also to earlier state saturation, which directly contributes to computational savings under our early-exit framework.
%
%
%
\section{Additional Results}
\label{sec:additional_results}
 
\subsection{Architecture Hyperparameters}
\label{subsec:hyperparams}
 
Table~\ref{tab:hyperparams} lists the key architectural hyperparameters of
RAVEN. The antenna mixer is deliberately narrow (64 dims, 8 heads) so that it
adds negligible GMACs on top of the channel SSMs; the Mamba state dimension of
16 keeps per-RX encoders lightweight; and the $1{\times}1$ Conv1D projection
maps chirp features to a $32{\times}56$ BEV grid before the detection and
segmentation decoders.
 
\begin{table}[h]
\centering
\small
\setlength{\tabcolsep}{3pt}
\renewcommand{\arraystretch}{1.1}
\begin{tabularx}{\columnwidth}{@{}lX@{}}
\toprule
\textbf{Component} & \textbf{Configuration} \\
\midrule
Antenna Mixer      & Dim 64, 8 heads, expansion $4{\times}$, init $\sim\!\mathcal{N}(0,1)$ \\
SSM (Mamba)        & State dim 16, conv kernel 4, expansion 2 \\
Spatial Proj.      & $1{\times}1$ Conv1D $\to$ 1792 ch.\ (grid $32{\times}56$) \\
\bottomrule
\end{tabularx}
\caption{\textbf{RAVEN architectural hyperparameters.}}
\label{tab:hyperparams}
\end{table}
 
To quantify the impact of compressing per-RX ADC samples to a single token, we
test RAVEN variants where the fast-time SSM condenses each RX channel into
$K \in \{1, 4, 8, 16\}$ tokens before the cross-antenna mixer.
Table~\ref{tab:token_ablation} shows that the marginal gain from $K{=}1$ to
$K{=}16$ is only $0.3\%$ F1, confirming that fast-time samples are highly
compressible and that a single token captures the essential range/phase
information needed downstream, validating our design choice.
 
\begin{table}[h]
\centering
\small
\setlength{\tabcolsep}{6pt}
\begin{tabular}{@{}lcccc@{}}
\toprule
\textbf{Tokens per RX ($K$)} & \textbf{1 (RAVEN)} & \textbf{4} & \textbf{8} & \textbf{16} \\
\midrule
\textbf{F1 (Detection)}       & 0.934              & 0.936      & 0.936      & 0.937 \\
\bottomrule
\end{tabular}
\caption{\textbf{Fast-time token compression ablation on RADIal.}
Marginal gains from $K{=}1$ to $K{=}16$ confirm that a single token per RX
channel sufficiently captures range and phase information.}
\label{tab:token_ablation}
\end{table}
 
\subsection{Early-Exit Decision Rule: Cosine Similarity vs.\ Entropy}
\label{subsec:early_exit_ablation}
 
\begin{figure}[h]
    \centering
    \includegraphics[width=\linewidth]{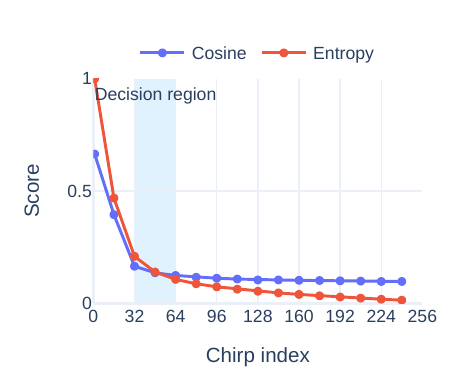}
    \caption{\textbf{Cosine distance vs.\ entropy as chirp-stopping signals.}
    Cosine similarity (blue) produces a cleaner knee-point, enabling more
    consistent early-exit decisions than entropy (orange).}
    \label{fig:early_exit_analysis}
\end{figure}

We compare two chirp-stopping criteria: (i)~minimum cosine similarity between
the new chirp latent state and all prior states (our default), and
(ii)~entropy of the chirp-state distribution. Although entropy produces a
smoother signal, cosine similarity yields better validation performance
(+0.85\% mAP, +0.67\% mIoU) at similar compute, as shown in
Table~\ref{tab:early_exit_rules} and Figure~\ref{fig:early_exit_analysis}.
The cosine rule directly measures the novelty of each new chirp in the latent
space, providing a more reliable and interpretable stopping condition.
 
\begin{table}[h]
\centering
\small
\setlength{\tabcolsep}{5pt}
\begin{tabular}{@{}lcccc@{}}
\toprule
\textbf{Stopping Rule} & \textbf{mAP} & \textbf{mAR} & \textbf{F1} & \textbf{mIoU} \\
\midrule
Cosine (Ours)  & \textbf{94.5} & \textbf{95.1} & \textbf{94.8} & \textbf{89.5} \\
Entropy        & 93.6          & 94.0          & 93.8          & 88.8 \\
\bottomrule
\end{tabular}
\caption{\textbf{Early-exit decision rule comparison on RADIal} (all metrics
in \%). Cosine similarity outperforms entropy on every metric.}
\label{tab:early_exit_rules}
\end{table}
 
\subsection{Adaptive Chirp Selection vs.\ Scene Velocity}
\label{subsec:vel_analysis}
 
\begin{figure}[h]
    \centering
    \includegraphics[height=3.2cm]{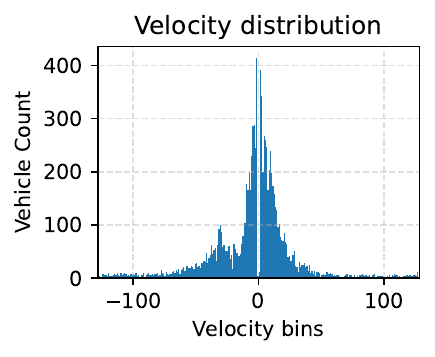}
    \hfill
    \includegraphics[height=3.2cm]{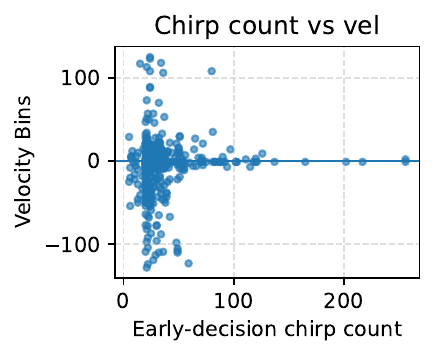}
    \caption{\textbf{Velocity distribution and adaptive chirp count.}
    (Left) Velocity histogram of annotated objects in RADIal.
    (Right) Scatter plot of per-frame selected chirp count vs.\ object velocity.
    The absence of correlation confirms that adaptive stopping is
    stability-driven, not velocity-driven.}
    \label{fig:vel_distro}
\end{figure}

Although static scenes nominally require less Doppler resolution, multiple
chirps are still needed to form the virtual MIMO aperture for angular cues;
fewer chirps shrink the virtual array and degrade spatial localization.
Figure~\ref{fig:vel_distro} shows no correlation between selected chirp count
and object velocity, confirming that our adaptive stopping rule is driven by
\emph{prediction stability} in the latent space rather than scene motion.
 
\subsection{Multi-Task vs.\ Task-Specific Performance}
\label{subsec:multitask}
 
Joint training does not introduce gradient interference. RAVEN trained jointly
outperforms task-specific single-head baselines on both objectives: detection
(0.95 vs.\ 0.93 mAP) and segmentation (90.2\% vs.\ 90.1\% mIoU). We attribute
this to the shared chirp-SSM backbone learning complementary spatial features
that benefit both heads simultaneously.
 
 


\end{document}